\documentclass[prd, aps, nofootinbib, preprintnumbers, showpacs, superscriptaddress,twocolumn]{revtex4}

\usepackage{graphicx}
\usepackage{bm}
\usepackage{amsmath}
\usepackage{amsfonts}
\usepackage{times}
\usepackage{color}

\newcommand{\Ref}[1]{(\ref{#1})}
 
\def\be{\begin{equation}}
\def\ee{\end{equation}}
\def\bea{\begin{eqnarray}}
\def\eea{\end{eqnarray}}
\def\l{\ell}
\def\EOB{\rm EOB}
\def\DIS{\rm DIS}
\def\NR{\rm NR}
\def\pole{\rm pole}
\def\i{\rm i}

\begin{document}

\title{Faithful Effective-One-Body waveforms of \\
equal-mass coalescing black-hole binaries}

\author{Thibault \surname{Damour}}
\affiliation{Institut des Hautes Etudes 
Scientifiques, 91440 Bures-sur-Yvette, France}
\affiliation{ICRANet, 65122 Pescara, Italy} 
\author{Alessandro \surname{Nagar}}
\affiliation{Institut des Hautes Etudes 
Scientifiques, 91440 Bures-sur-Yvette, France}
\affiliation{INFN, Sez. di Torino, Via P.~Giuria 1, Torino, Italy}
\affiliation{ICRANet, 65122 Pescara, Italy}
\author{Ernst Nils Dorband}
\affiliation{Max-Planck-Institut f\"ur Gravitationsphysik,
  Albert-Einstein-Institut, Potsdam-Golm,Germany}
\author{Denis Pollney}
\affiliation{Max-Planck-Institut f\"ur Gravitationsphysik,
  Albert-Einstein-Institut, Potsdam-Golm,Germany}
\author{Luciano Rezzolla}
\affiliation{Max-Planck-Institut f\"ur Gravitationsphysik,
  Albert-Einstein-Institut, Potsdam-Golm,Germany}
\affiliation{Department of Physics and Astronomy,
  Louisiana State University,
  Baton Rouge, LA, USA}

\begin{abstract}
  We compare a recently derived, {\it resummed} high post-Newtonian accuracy
  Effective-One-Body (EOB) quadrupolar waveform to
  the results of a numerical simulation of the {\it inspiral and merger} 
  of an {\it equal-mass} black hole binary.  We find a remarkable
  agreement, both in phase and in amplitude, with a maximal dephasing
  which can be reduced below $\pm 0.005$ gravitational-wave (GW) cycles
  over 12 GW cycles corresponding to the end of the inspiral, the plunge, 
  the merger and the beginning of the ringdown. This level of agreement 
  is shown for two different values of the effective fourth post-Newtonian 
  parameter $a_5$, and for  corresponding, appropriately ``flexed'' values 
  of the radiation-reaction resummation
  parameter $v_{\rm pole}$. In addition, our resummed EOB amplitude
  agrees to better than the $1\%$ level with the numerical-relativity one
  up to the late inspiral. 
  These results, together with other recent work on the
  EOB-numerical-relativity comparison, confirm the  ability of the EOB 
  formalism to accurately capture the general relativistic waveforms.
\end{abstract}

\date{\today}


\maketitle

\section{Introduction}
\label{sec:intro}
The gravitational-wave (GW) signals emitted by coalescing black hole binaries are  
among the most promising targets for the currently operating network of ground-based 
detectors  GEO/LIGO/VIRGO. The most useful part of the waveform for 
detection comes from the most relativistic part of the dynamics,
around the coalescence, i.e. the last few cycles of the adiabatic
inspiral, the plunge and the merger.
It is crucial for GW detection purposes to have available  
a large bank of  ``{\it templates}'' that accurately represent the GW
signals radiated by these binaries. 
The construction of {\it faithful}\footnote{Following the terminology 
of~\cite{Damour:1997ub}, we recall that ``effectual templates'' are templates 
exhibiting large overlaps with an exact signal after maximizing over all 
(kinematical and dynamical) parameters, while ``faithful'' ones are so 
``close'' to an exact one that they have large overlaps for values of 
the dynamical parameters which are very close to the real ones (``small biases'').}
GW templates for coalescing binaries comprising spinning black
holes (with arbitrary masses $m_1$, $m_2$ and spins ${\bf S}_1$, ${\bf S}_2$) 
is a non trivial task. In view of the multi-dimensionality of the 
corresponding parameter space,  state-of-the-art numerical simulations 
cannot densely sample this parameter space. It is therefore urgent
to devise {\it analytical} methods for computing (as a function of
the physical parameters $m_1$, $m_2$, ${\bf S}_1$, ${\bf S}_2$) the
corresponding GW waveforms. 
Here we continue the program of constructing, within the Effective-One-Body
(EOB) method~\cite{Buonanno:1998gg,Buonanno:2000ef,Damour:2000we,gr-qc/0103018}
high-accuracy analytic waveforms describing the GW signal emitted
by inspiralling and merging binary black holes with arbitrary masses and spins. 
The EOB method was the first to provide estimates of the complete waveform 
(covering inspiral, plunge, merger and ring-down)
of a coalescing black hole binary, both for non-spinning systems
\cite{Buonanno:2000ef}, and for spinning ones~\cite{Buonanno:2005xu}.

Numerical Relativity (NR) recently succeeded in giving us
access to reliable information 
about the dynamics and radiation of binary black hole coalescences
\cite{gr-qc/0507014,Campanelli:2005dd,Campanelli:2006uy,Baker:2006yw,Baker:2006vn,Baker:2007fb,Gonzalez:2006md,
arXiv:0706.0740,Koppitz:2007ev,arXiv:0708.3999,arXiv:0710.3345,arXiv:0710.0158}.
This opens the possibility of comparing the EOB predictions to NR results.

The comparison between the EOB approach and NR results has been recently
initiated in several works~\cite{gr-qc/0610122,arXiv:0704.1964,arXiv:0704.3550,arXiv:0705.2519,
arXiv:0706.3732,arXiv:0711.2628}. These recent comparisons have been done
using two different versions of EOB waveforms. The works of Buonanno et al.
\cite{gr-qc/0610122,arXiv:0704.1964,arXiv:0706.3732} used a {\it restricted
waveform}, as proposed in the first EOB paper~\cite{Buonanno:2000ef}, 
but with an improved matching to the ringdown (similar to the one used 
in~\cite{Damour:2006tr})   making use of three quasi-normal
modes. By contrast, the recent works of Damour and Nagar~\cite{arXiv:0705.2519,
arXiv:0711.2628} use a new, {\it resummed} high post-Newtonian (PN) 
accuracy~\footnote{This high PN accuracy can be called $3^{+2}$-PN because it
includes not only the known comparable-mass 3~PN waveform corrections,
but also the test-mass limit of the 4~PN and 5~PN waveform amplitude
corrections. See~\cite{arXiv:0711.2628} for details and references.} 
EOB  quadrupolar waveform. This improved EOB waveform has been shown to exhibit a remarkable
agreement, both in phase and in amplitude, with NR waveforms in two separate
physical situations:  (i) inspiral and coalescence of small-mass-ratio 
(non-spinning) systems~\cite{arXiv:0705.2519} (comparing it to
waveforms computed by means of numerical simulations of 
test particles, with an added radiation-reaction force, moving
in black-hole backgrounds~\cite{Nagar:2006xv})  and (ii) inspiral (up to
a limiting GW frequency $\sim 0.14/M$) of an equal-mass (non-spinning) system
\cite{arXiv:0711.2628} (comparing it to  recently published results of
a high-accuracy inspiral simulation~\cite{arXiv:0710.0158}). 

The present paper is a continuation of the general program of 
constructing, within the Effective-One-Body (EOB)
approach, high accuracy, faithful analytic waveforms describing the
gravitational wave signal emitted by inspiralling and coalescing binary
black holes. Here we shall consider the {\it coalescence signal} emitted
by a non-spinning {\it equal-mass} binary black-hole system. We shall
compare the phase and the amplitude of the new {\it resummed}
$3^{+2}$~PN-accurate EOB  quadrupolar waveform
of~\cite{arXiv:0705.2519,arXiv:0711.2628} to a numerical relativity 
simulation of a coalescing black hole binary performed
at the Albert Einstein Institute (AEI).

This comparison will confirm the ability of the EOB approach to provide 
accurate analytical representations of NR waveforms.
We note that the recent work~\cite{arXiv:0706.3732} had already shown the ability
of the analytically less accurate {\it restricted}~\footnote{Here 
``restricted'' refers to a waveform $h\propto \Omega^{2/3}e^{-2\i\Phi}$.} 
EOB waveforms to provide rather accurate approximations to NASA-Goddard NR 
coalescence waveforms for several different mass ratios ($m_1/m_2 = 1,3/2,2$
and 4). More precisely, Ref.~\cite{arXiv:0706.3732} found, in the
equal mass case, an EOB/NR dephasing of $\sim \pm 0.03$ GW cycles over 
15 GW cycles. Concerning the amplitude, the latter reference does not 
quantify the restricted EOB/NR difference, but one can read 
from Fig.~21 of~\cite{arXiv:0710.0158} that the  
difference between the restricted PN (or EOB) amplitude 
and the Caltech-Cornell inspiral NR one is~$\sim 7\%$.
By contrast, the present paper will show that the new, resummed waveform 
exhibits a significantly smaller dephasing $\sim \pm 0.005$ GW cycles over 
12 GW cycles, and, most remarkably, exhibits an excellent agreement in 
amplitude, both during the inspiral (where it is better than the $1\%$ level)  
and the ring-down. This good result is obtained by making use 
(as proposed in several previous works~\cite{gr-qc/0103018,gr-qc/0204011,
gr-qc/0211041,arXiv:0706.3732,arXiv:0711.2628}) of the natural {\it flexibility}
of the EOB approach.

An alternative approach to the construction of analytical templates
to model (non-spinning) coalescing binary black holes with arbitrary mass ratios
has been recently proposed in Refs.~\cite{Ajith_et_al:2007a,Ajith_et_al:2007b}.

This paper is organized as follows: In Sec.~II we briefly describe the 
numerical simulation, whose  results we use in the 
following. In Sec.~III we spell out the features of the EOB waveform that we shall
use. The main section is Sec.~IV where we compare the new, resummed EOB
waveform to NR data. We also include a comparison where we use the
less accurate ``restricted'' EOB waveform, and simpler QNM-matching, used in
some of the previous EOB works~\cite{Buonanno:2000ef,gr-qc/0610122,arXiv:0704.1964,arXiv:0706.3732}.
The paper ends with some conclusions.

\section{Brief description of the numerical simulation}
\label{sec:nr}

The numerical simulations have been carried out with the
\texttt{Ccatie} code~\cite{Pollney:2007ss}, a three-dimensional
finite-differencing code developed at the Albert Einstein Institute
and at the Center for Computation and Technology (CCT) of the
Louisiana State University. The code is based on the \texttt{Cactus}
Computational Toolkit~\cite{Goodale02a} for the solution of the
Einstein equations in a finite-size domain covered with a Cartesian
rectangular grid. The main and new features of the code have been
recently discussed in Ref.~\cite{Pollney:2007ss}, and we here briefly
recall the most important ones only.

The Einstein equations are formulated as an initial-value problem via
a conformal and traceless ``$3+1$'' decomposition. 
The spacetime geometry is decomposed into: (i) the 3-metric of  
spacelike slices, (ii) the extrinsic curvature of those slices, and 
(iii) the lapse and shift. See~\cite{Pollney:2007ss} for the explicit form
of the equations. The lapse function is evolved using the
``$1+\log$'' slicing condition~\cite{Bona:1994dr}, while
the shift is evolved using the hyperbolic $\tilde{\Gamma}$-driver
condition discussed in Ref.~\cite{Alcubierre:2002kk}, but with the
difference that advection terms have been added following the
experience of~\cite{Campanelli:2005dd,Baker:2005vv}, and are required
for correct advection of the punctures in ``moving-puncture''
evolutions.

Spatial differentiation of the evolution variables is performed via
straightforward finite-differencing using fourth-order accurate
centered stencils for all but the advection terms for each variable,
which are instead upwinded in the direction of the
shift. Vertex-centered adaptive mesh-refinement is employed using
nested grids via the \texttt{Carpet}
infrastructure~\cite{Schnetter-etal-03b}, with a $2:1$ refinement for
successive grid levels, and the highest resolution concentrated in the
neighborhood of the individual horizons. Individual apparent horizons
are located every few time steps during the time
evolution~\cite{Thornburg2003}, which is obtained via a
``method-of-lines'' and with a fourth-order accurate Runge-Kutta time
integrator.

The simulations were performed on a domain with outer boundaries
located at~\footnote{We denote by $M_c$ the 
internal length and mass units used in the code (with $G=c=1$). Beware that
$M_c$ slightly differs from $M=m_1+m_2$ (see below).} $768M_c$, 
and a grid structure consists of nine mesh-refinement levels, 
the finest of which has a spatial resolution of $h=0.02M_c$.
Simulations with lower resolution (i.e., with $h=0.024M_c$ and
$h=0.03M_c$) have also been carried out to validate the consistency of
the results. An important feature of the \texttt{Ccatie} code is the
possibility of employing two distinct methods for the calculation of
the gravitational radiation produced.  The first method uses the
Newman-Penrose curvature scalar $\psi_4$, with respect to a suitable
frame at the extraction radius. 
An alternative method measures the metric of the
numerically generated spacetime against a fixed background at the
extraction radius, and determines the gauge-invariant Regge-Wheeler-Zerilli-Moncrief
functions (see Ref.~\cite{gr-qc/0502064} for a review and references). 
Both methods have been systematically studied in Ref.~\cite{Pollney:2007ss}, 
where they were also compared
and shown to yield essentially identical results, both in terms of
their asymptotic scaling properties (e.g., the
peeling-theorem), and in terms of the polarization amplitudes $h_{+}$
and $h_{\times}$. The analysis carried out here used as basic NR data 
the gauge-invariant (Zerilli-Moncrief) metric perturbations. These
were extracted on (NR) coordinate 2-spheres with (NR) 
coordinate radii $R_{\NR}=60M_c$ up to $R_{\NR}=120M_c$, 
with a separation of $10M_c$ between two adjacent observers.
The analysis carried out below uses, as approximate asymptotic amplitude, 
the metric perturbation extracted at $R_{\NR}=120M_c$. 
%
%
\begin{table*}[t]
\caption{\label{tab:table1}Initial ADM mass (scaled by $M=m_1+m_2$)
   and angular momentum of the spacetime (scaled by $M^2$); final mass
   (scaled by $M$) and dimensionless spin parameter $j_{\rm f}=J_{\rm
   f}/M_{\rm f}^2$ of the merged black hole; dominant (quasi-normal-mode)
   complex frequency of the ringdown; for two different grid spacings $h$.}
\begin{center}
  \begin{ruledtabular}
  \begin{tabular}{ccccccccc}
    $h/M$ & $M_{\mathrm{ADM}}/M$ & $J_{\rm ADM}/M^2$ & $M^{\rm hor}_{\rm f}/M$ & $j^{\rm
    hor}_{\rm f}$ & $M^{\rm ring}_{\rm f}/M$ & $j^{\rm ring}_{\rm f}$& $M\sigma_{2220}^+$  & \\ 
    \hline \hline
    0.024 & 0.990484 & 0.991803 & 0.951531 & 0.687142 & -- & -- & -- \\    
    0.020 & 0.990484 & 0.991803 & 0.951611 & 0.686916 &0.959165 & 0.684639 &
    0.085475 + {\i} 0.551040 \\  
  \end{tabular}
\end{ruledtabular}
\end{center}
\label{tab:MassAndSpin}
\end{table*}%

The initial data for the black-hole binary are obtained by a
Brill-Lindquist~\cite{Brill:1963yv} construction, where the additional 
asymptotically flat end of each wormhole is compactified into a single 
point, the so called \emph{puncture}~\cite{Brandt:1997tf}. 
This approach explicitly uses the Bowen-York extrinsic curvature 
and solves the Hamiltonian constraint equation numerically 
(as detailed in Ref.~\cite{Ansorg:2004ds}), after having 
chosen the free parameters for the puncture initial data.
Quasi-circularity of the initial
orbit can then be obtained by specifying the puncture parameters in
terms of an effective-potential method~\cite{Cook:1994va} as 
discussed in~\cite{Pollney:2007ss}. 
However, the assumption of ``quasi-circularity'' 
(in the sense of~\cite{Cook:1994va}) at the 
(rather small) initial separations frequently used in
numerical-relativity simulations has the drawback of introducing a
small but nonzero amount of eccentricity. To compensate for, or reduce,
this effect, other approaches have been suggested recently. One of
these is based on an iterative minimization procedure where,
throughout a series of simulations with slightly different initial
black hole configurations, the eccentricity is measured and
minimized~\cite{Pfeiffer:2007yz}. A simpler and rather effective
approach has been proposed in Ref.~\cite{Husa:2007rh}, 
and consists of specifying the initial puncture-parameters as the
end-state of a binary system whose evolution is determined, starting
from a large separation, via the solution of the Taylor-expanded 
3~PN-accurate equations of motion~\cite{Damour:2001bu,Blanchet:2004ek,Buonanno:2005xu}.

We have here essentially followed this latter prescription and
considered, in particular, the initial data denoted by \emph{E11} in
Table I of~\cite{Husa:2007rh}, that have been shown there to reduce
the eccentricity to $e < 0.002$.  More specifically, our initial black
holes have a coordinate distance $D=11M_c$, momenta in the radial and
tangential directions of $P_r = -7.09412\times10^{-4}M_c$ and $P_t =
0.0900993M_c$, and a puncture mass-parameter of $0.487035M_c$, leading
to initial individual black-hole masses $m_1 = m_2 = 0.499821M_c$, and
thus a total mass of the binary system
$M=m_1+m_2=0.999642M_c$. Overall, the simulation covers about $\sim
1600\,M$ of the final evolution of the binary, thus comprising $8$
orbits and about $16$ GW cycles.

The mass and spin of the final black hole have been computed through
two different methods yielding, however, very similar results: 
(i) by using the isolated/dynamical horizon 
formalism~\cite{Dreyer:2002mx, Ashtekar:2003hk}, where a proper
rotational Killing vector is searched on the final apparent horizon to
measure the spin, and the horizon area is used for computing the black
hole mass (see Sec.~IV~D of Ref.~\cite{Pollney:2007ss} for
details); (ii) by performing a fit of the dominant
quasi-normal mode\footnote{In the notation introduced
in Sec.~\ref{sec:eob} below, the dominant mode corresponds to the labels
$(\pm,\l,\l',m,n)=(+,2,2,2,0)$.} of the {\it complex} ringdown waveform. 
This fit was performed by a non-linear least-squares Gauss-Newton
method, using $\exp(-\sigma t + \rho)$  as a parameter--dependent 
template (with two {\it complex} parameters ($\sigma,\rho$)), 
and an appropriate time interval during the ringdown 
(chosen by minimizing the post-fit residual).
[For a discussion of methods for QNM fitting see 
Refs.~\cite{Dorband:2006gg,Berti:2007dg,Berti:2007fi}].
Then, from the best-fit  value of $\sigma$ (i.e., the QNM dominant
complex frequency  $\sigma^+_{2220}$), we computed  the values of the 
mass and dimensionless spin parameters of the final
black hole by using the interpolating fits given in Appendix~E of 
Ref.~\cite{gr-qc/0512160}.  The results of these two methods are 
denoted as $(M^{\rm hor},j^{\rm hor})$ and $(M^{\rm ring},j^{\rm ring})$, 
respectively. 

The most relevant properties of the binary system are summarized 
in Table~\ref{tab:table1}. The difference (which is $\lesssim 1\%$) 
between the quoted values of the final black hole parameters might 
come, in part, from inaccuracies in the interpolating fits of 
Ref.~\cite{gr-qc/0512160}. In the following we will use, in our
EOB-matching procedure, the ringdown--fitted black hole parameters
$(M^{\rm ring},j^{\rm ring})$ (so that the dominant complex 
frequency will be guaranteed to have the best possible value).

\section{Effective-One-Body (EOB) method and waveform}
\label{sec:eob}

We shall not review here in detail the EOB method
\cite{Buonanno:1998gg,Buonanno:2000ef,Damour:2000we,gr-qc/0103018}, 
which has been described in several
recent publications, notably Refs.~\cite{arXiv:0706.3732,arXiv:0711.2628}.
We shall only indicate the EOB elements  that are crucial for the present study.
For detailed definitions of the EOB ingredients we refer to the recent 
paper~\cite{arXiv:0711.2628} that we follow, except when otherwise 
indicated below.

Before entering the details of our EOB implementation, let us recall that
Ref.~\cite{arXiv:0711.2628} proposed a methodology for improving the
waveform implementation of the EOB philosophy based on understanding, element by element,
the physics behind each feature of the waveform, and on systematically comparing  various
EOB-based waveforms with ``exact'' waveforms obtained by numerical relativity approaches.
The first step of the methodology consisted in studying the small-mass-ratio limit,
$ \nu\equiv m_1 m_2/M^2 \ll 1$, in which one can use the well controllable
``laboratory'' of numerical simulations of 
test particles (with an added radiation-reaction force) moving
in black hole backgrounds. Historically, this ``laboratory'' has
been important in understanding/discovering several key features of
GW emission near black holes. A notable example of this being the work of
Davis, Ruffini and Tiomno ~\cite{Davis:1972ud} which discovered the transition between
the plunge signal and a ringing tail when a particle falls into a black hole.
The recent study of inspiralling and merging small-mass-ratio systems~\cite{arXiv:0705.2519}
led to introducing (and testing) the following improvements in EOB dynamics and waveforms:
(i) an improved analytical expression for the ($(\ell,m) =(2,2)$ even-parity 
Zerilli-Moncrief) waveform $\Psi_{22}^{(\rm e)}$  which includes a resummation 
of the tail effects, and a $3^{+2}$~PN-accurate ``non-linear'' amplitude 
correction, (ii) the inclusion of non-quasi-circular corrections
to the waveform, (iii)  the inclusion of non-quasi-circular corrections
to radiation reaction, and (iv) an improved treatment of the matching between the plunge
and ring-down waveforms which takes into account a new understanding of the
importance of the number of quasi-normal-modes (QNMs), the sign of their frequencies, and the 
length of the interval on which the matching is done. The resulting
improved implementation (when $ \nu \ll 1$) of the EOB approach yielded very faithful
waveforms whose amplitude and phase agreed remarkably well with the ``exact'' ones:
in particular, the EOB phasing differed from the ``exact'' one by less than
$\pm 1.1 \%$ of a cycle over the whole process. 

The program initiated in~\cite{arXiv:0705.2519} was pursued in~\cite{arXiv:0711.2628}
where the comparable-mass version of the improved, resummed $3^{+2}$-PN 
accurate waveform was compared with the recently published inspiral simulation of the
Caltech-Cornell group~\cite{arXiv:0710.0158}. It was found that, by exploiting 
the combined flexibility in $a_5$ and $v_{\rm pole}$, one could 
 reach a remarkable phase
agreement, better than $0.001$ GW cycles over 30 GW cycles.
 Here, we shall similarly exploit
the flexibility in $a_5$ and $v_{\rm pole}$ 
to best fit the AEI merger waveform.

Let us recall that the EOB approach is a {\it non-perturbatively resummed} 
analytic technique which consists of several different elements:

\begin{itemize}

\item a {\it Hamiltonian} $H_{\rm real}$ describing the conservative part
of the relative two-body dynamics. The key ingredient of this Hamiltonian 
(defined in Eqs. (13) and (14) of~\cite{arXiv:0711.2628}) is the ``radial potential''
$A(r)$.\footnote{Except when said otherwise, we henceforth systematically scale
dimensionful quantities by means of the total rest mass
$M\equiv m_1 + m_2$ of the binary system. For instance, we use the 
dimensionless EOB radial coordinate $r\equiv R_{\rm EOB}/M$, with $G=1$. Note also that
$\nu \equiv \mu/M$ with $\mu \equiv m_1 m_2/M$.}
 This radial potential is defined, at n-Post-Newtonian (PN) order, as the (1,n) 
Pad\'e resummation~\cite{Damour:2000we} of its Taylor (i.e. usual PN) expansion
(written in Eq. (15) of~\cite{arXiv:0711.2628}). 
 
\item a {\it radiation reaction force} ${\cal F}_{\varphi}$ (denoted $\hat{\cal F}_{\varphi}$ 
after its rescaling by $1/\mu$), which is defined as a Pad\'e resummation
\cite{Damour:1997ub} of its Taylor expansion. 
See Eq.~(17) of~\cite{arXiv:0711.2628} where
$f_{\rm DIS}$ is the $P^{4}_{4}$ 
Pad\'e resummation of $(1-v/v_{\rm  pole})\hat{F}^{\rm Taylor}(v;\nu)$.
The coefficients of $\hat{F}^{\rm Taylor}$ in Eq.~(18)
of~\cite{arXiv:0711.2628} have been derived in 
Refs.~\cite{Blanchet:1995ez,gr-qc/0105098,gr-qc/0406012,Blanchet:2005tk,Blanchet:2001ax,Tagoshi:1994sm}.
We shall also consider at the end, following Ref.~\cite{arXiv:0705.2519},
the possibility of modifying $\cal F_{\varphi}$ by a 
{\it  non-quasi-circular} correcting factor, Eq.~\eqref{rr_nqc}.

\item {\it improved ``post-post-circular''} dynamical initial data 
(positions and momenta) as advocated in Sec.~III~B of~\cite{arXiv:0711.2628}.
To explain the improved construction of initial
data let us introduce a formal book-keeping parameter $\varepsilon$
(to be set to 1 at the end) in front of the radiation reaction 
$\hat{\cal F}_\varphi$ in the EOB equations of motion. One
can then show that the quasi-circular inspiralling solution of 
the EOB equations of motion formally satisfies
\begin{align}
p_{\varphi}& = j_0(r) + \varepsilon^2 j_2(r) + O(\varepsilon^4), \\
p_{r_*}    & =\varepsilon\pi_1(r) + \varepsilon^3\pi_3(r) + O(\varepsilon^5).
\end{align}
Here, $j_0(r)$ is the usual {\it circular} approximation to
the inspiralling angular momentum as explicitly given by
Eq.~(4.5) of~\cite{Buonanno:2000ef}, while the order $\varepsilon$ 
(``post-circular'') term $\pi_1(r)$ is obtained by: 
(i) inserting the circular approximation 
$p_\varphi=j_0(r)$ on the left-hand side (l.h.s) of Eq.~(10)
of~\cite{arXiv:0704.3550}, (ii) using the chain rule
$dj_0(r)/dt=(dj_0(r)/dr)(dr/dt)$, (iii) replacing $dr/dt$ by
the right-hand side (r.h.s) of Eq.~(9) of~\cite{arXiv:0704.3550}
and (iv) solving for $p_{r_*}$ at the first order in $\varepsilon$.
This leads to an explicit result of the form (using the notation
defined in Ref.~\cite{arXiv:0704.3550})
\begin{align}
\label{p_adiab}
\varepsilon \pi_1(r)
= \left[\nu\hat{H}\hat{H}_{\rm eff}\left(\dfrac{B}{A}\right)^{1/2}\left(\dfrac{dj_0}{dr}\right)^{-1}\hat{\cal
    F}_\varphi\right]_0,
\end{align} 
where the subscript $0$ indicates that the r.h.s. is evaluated
at the leading circular approximation $\varepsilon\to 0$.
The post-circular EOB approximation $(j_0,\pi_1)$ was introduced
in Ref.~\cite{Buonanno:2000ef} and then used in most of the subsequent
EOB papers~\cite{Buonanno:2005xu,gr-qc/0610122,arXiv:0704.3550,
arXiv:0706.3732,arXiv:0706.3732,Nagar:2006xv}.
The {\it post-post-circular} approximation (order $\varepsilon^2$), 
introduced in Ref.~\cite{arXiv:0711.2628} and used here, consists of: (i) formally 
solving Eq.~(11) of~\cite{arXiv:0704.3550} with respect to the explicit
$p_{\varphi}^2$ appearing on the r.h.s., (ii) replacing $p_{r_*}$ by
its post-circular approximation~(\ref{p_adiab}), (iii) using the 
chain rule $d\pi_1(r)/dt = (d\pi_1(r)/dr)(dr/dt)$, and (iv) replacing
$dr/dt$ in terms of $\pi_1$ (to leading order) by using Eq.~(9) 
of~\cite{arXiv:0704.3550}. The result yields an explicit expression
of the type $p_\varphi^2 \simeq j_0^2(r)[1 + \varepsilon^2 k_2(r)]$ of which
one finally takes the square root. In principle, this procedure can
be iterated to get initial data at any order in $\varepsilon$.
We found that the post-post-circular initial data $(j_0\sqrt{1+ \varepsilon^2k_2},\pi_1)$
are sufficient to lead to negligible eccentricity when 
starting the integration of the EOB equations of motion 
at radius $r=15$.

\item an improved, resummed  {\it ``inspiral-plus-plunge''} (hereafter abbreviated as
{\it ``insplunge''}) {\it  waveform}\footnote{Here, as before, we work with a
{\it metric-level} (``$h$''), rather than {\it curvature-level} (``$\psi_4$''),
waveform. However, we normalize here this metric waveform in the same 
``Zerilli-Moncrief'' way as in the  test-mass work~\cite{arXiv:0705.2519}.
This differs simply by a numerical factor from both the usual tensor-spherical
harmonics $(\ell,m)$ metric amplitude $h_{\ell  m}$ and the related metric
variables $Q_{\l m}^{+,\times}$ extracted from the NR 
evolution~\cite{Pollney:2007ss}:
$R h_{\l  m}=\sqrt{(\l+2)(\ell+1)\ell(\ell-1)} \left( \Psi_{\l m}^{(\rm e)} 
+ \i\Psi^{(\rm o)}_{\l m} \right)
=\frac{1}{\sqrt{2}}\left( Q^+_{\l m} -{\i}\int_{-\infty}^t Q^\times_{\l m}(t')dt'\right) $ } of the form

\be 
 \label{hinsplunge}
\left(\dfrac{ c^2}{GM}\right) \Psi_{22}^{\rm insplunge}(t)=-4\sqrt{\dfrac{\pi}{30}}\nu
     (r_{\omega}\Omega)^2 f_{22}^{\rm NQC} F_{22} e^{-2{\i}\Phi} \ ,
\ee
where $\Phi(t)$ is the EOB orbital phase, $\Omega=\dot{\Phi}$ is the
EOB orbital frequency, $r_{\omega}\equiv r\psi^{1/3}$ is a 
modified EOB radius, with $\psi$ being defined in Eq.~(22) of 
Ref.~\cite{Damour:2006tr}. The factor $F_{22}$ is a resummed, 
$3^{+2}$-PN-accurate complex amplitude correction valid during 
the (adiabatic) inspiral, and $f_{22}^{\rm NQC}$ is an extra complex  correcting factor, aimed at
taking care (in an effective way) of  various {\it non quasi-circular} (NQC) effects
during the plunge. $F_{22}$ is defined in Eqs. (5)-(11) of~\cite{arXiv:0711.2628},
with $f_{22}$ being the (3,2) Pad\'e resummation of  
$f_{22}^{\rm  Taylor}$ [see also Ref~\cite{Kidder:2007rt} for an 
independent derivation of the nonresummed, 3~PN-accurate $(2,2)$ waveform].

\item a {\it ringdown waveform}
\be
\label{hringdown}
\Psi^{\rm ringdown}_{22}(t) = \sum_N C_N^{+} e^{-\sigma_N^{+} t}+ 
\sum_N C_N^{-} e^{-\sigma_N^{-} t} \ ,
\ee
where the label $N$ actually refers to a set of indices $(\l, \l', m, n)$, with
$(\l,m) = (2,2)$ being the Schwarzschild-background multipolarity degrees
of the considered (Zerilli-Moncrief-type) waveform $\Psi_{\l m} \sim h_{\l m}$, 
with $n=0,1,2,...$ being the ``overtone number'' of the considered 
Kerr-background Quasi-Normal Mode (QNM; $n=0$ denoting the fundamental mode),
and $\l'$ the degree of its associated spheroidal harmonics $S_{\l' m}(a \sigma, \theta)$.
In addition  $\sigma_N^{\pm}= \alpha_N^{\pm} \pm {\i}\omega_N^{\pm}$ refers to the
positive/negative  complex QNM frequencies  ($\alpha_N^{\pm} >0$ and
$\omega_N^{\pm}>0$  indicate the inverse damping time and the oscillation frequency of 
each mode respectively). The sum over $\l'$ comes from the fact that an
ordinary spherical harmonics $Y_{\l m}(\theta, \phi)$ (used as expansion basis 
to define $\Psi_{\l m}$) can be expanded in the spheroidal harmonics 
$S_{\l' m}(a \sigma, \theta) e^{{\i} m \phi}$ characterizing the angular
dependence of the Kerr-background QNMs~\cite{PT1973}.

\item an improved way of {\it matching} the inspiral-plus-plunge waveform to the ring-down one,
on a ($2p+1)$-tooth ``comb'' $(t_m - p \delta, t_m - (p-1) \delta,\ldots,
 t_m -  \delta, t_m , t_m + \delta, \ldots, t_m + p \delta)$, of total 
length $\Delta t=2p\delta$, which is centered around some ``matching'' time
$t_m$. Below we will fix the integer $p$ to the value $p=2$, 
corresponding to five matching points. 
 
 \item Finally, we define our complete EOB matched waveform 
       (from $t=- \infty$ to $t=+ \infty$) as

\begin{align}
\label{hmatched}
\Psi^{\rm EOB}_{22}(t)&\equiv \theta(t_m -t) \Psi^{\rm insplunge}_{22}(t)\nonumber\\
                      &+ \theta(t -t_m ) \Psi^{\rm ringdown}_{22}(t)
\end{align}
where $\theta(t)$ denotes Heaviside's step function. Note that,
if one wanted to have a
$C^{\infty}$ transition between the two waveforms one could replace $\theta(t -t_m)$ by one
of Laurent Schwartz's well-known smoothed step functions (or ``partitions of unity'')
 ${\theta_{\varepsilon}}((t-t_m)/(2p\delta))$.

\end{itemize}

Let us now state the specific choices made here for the various EOB ingredients just
recalled. Some of these choices correspond to various ways of ``flexing''  the
EOB formalism (in the sense of Ref.~\cite{gr-qc/0211041}).

\begin{itemize}

\item We ``flex'' the currently known 3~PN-accurate
EOB Hamiltonian~\cite{Damour:2000we,Damour:2001bu} by introducing an (effective)
4~PN Hamiltonian parameter $a_5$, parametrizing an additional contribution
$ +  a_5 \nu/r^5$ in the main EOB radial function $A(r)$. This parameter
has already been introduced (under varying notations) in several previous 
works~\cite{gr-qc/0103018,gr-qc/0204011,gr-qc/0211041,arXiv:0706.3732,arXiv:0711.2628}.

\item Similarly, the EOB radiation reaction force (defined by Eq. (17) of~\cite{arXiv:0711.2628})
is ``flexed'' by allowing the Pad\'e-resummation parameter $v_{\rm pole}$ to differ
from the ``standard'' value $v^{\rm DIS}_{\rm pole}(\nu)$ advocated in~\cite{Damour:1997ub}.

In addition, we shall also briefly explore another physically natural
flexibility in the radiation reaction, which was introduced (and shown to
be physically needed for faithfulness) in~\cite{arXiv:0705.2519}:
the multiplication of the radiation reaction by a non quasi-circular (NQC)
correction factor $f_{\rm RR}^{\rm NQC}$, see Eq.~\eqref{rr_nqc} below.

\item To define precisely the ``insplunge waveform'' \Ref{hinsplunge} we need to specify:

(i) the argument $x(t)$ used in the $f_{22}$ ``brick'' within $F_{22}$ (see Eq.(10) 
of~\cite{arXiv:0711.2628}). We shall use here $x= \Omega^{2/3}$ where $\Omega$
is the {\it EOB orbital frequency}.

(ii) the Pad\'e resummation of the Taylor
expansion $f_{22}^{\rm Taylor}$ of $f_{22}$. As in~\cite{arXiv:0711.2628}
we shall use a $P^3_2$ Pad\'e.

(iii) the definition of the non quasi-circular (NQC)
correction factor $f_{22}^{\rm NQC}$. To do this we follow the rationale
explained in~\cite{arXiv:0705.2519}. For convenience, we choose (as suggested in
footnote 9 of~\cite{arXiv:0705.2519}) a {\it factorized} complex NQC 
factor

\be
\label{eq:waveform}
f_{22}^{\rm NQC} = \left[ 1 + a \frac{ p_{r_*}^2}{(r\Omega)^2 + \epsilon}\right]
\exp \left(+ {\i} b \frac{p_{r_*}}{r\Omega}\right)  ,
\ee
in which $a$ (denoted $a'$ in the cited footnote)
affects only the modulus, and $b$ (alias $b'$) only the phase.
To ease some technical problems during the ring-down linked to the fact that 
$\Omega(t)$ tends exponentially towards zero as $t \to +\infty$ we have added
a (``cut-off'') constant $\epsilon$ to the first denominator $(r\Omega)^2$.
As discussed in~\cite{arXiv:0705.2519}, one can {\it a priori}  analytically determine 
a ``good'' value of the NQC-modulus parameter $a$ by requiring that the
modulus of the full EOB insplunge waveform \Ref{hinsplunge} be maximum at the
``EOB-light-ring'', i.e. when the EOB orbital frequency $\Omega$ reaches a maximum.
Ref.~\cite{arXiv:0705.2519} mentioned that, in the $\nu \ll 1$ limit, this requirement
implied $a = 1/2$ (when $\epsilon=0$). We found, by numerically exploring
the modulus of $\Psi_{22}^{\rm insplunge}(t)$, that the same value, 
$a=1/2$ (together with $\epsilon=0.12$), can be used in the case 
$\nu=1/4$ considered here. 
Concerning the NQC-phase parameter $b$ we simply choose $b=0$. 
[Note that the comparable-mass resummed EOB waveform of~\cite{arXiv:0711.2628} 
uses a refined estimate for the additional  phase $\delta_{22}$ of 
$ \Psi_{22}^{\rm insplunge}(t)$ compared to the one used in~\cite{arXiv:0705.2519}.]

\item Concerning the  choice of QNMs we recall that the discussion of the
physical excitation of QNMs in~\cite{arXiv:0705.2519} (see the summary in
Fig.~4 there) suggested that it is sufficient to use only {\it positive-frequency}
QNMs in the ringdown waveform \Ref{hringdown}. This is what we shall 
do here as well.

A new feature of the comparable-mass case (w.r.t. the small $\nu$ limit) is the ``mixing''
between various $\l'$ QNMs (with $\l' \neq \l$)
that can enter a given $ (\l,m)$ multipolar wave. This mixing is due to 
the ``$a \omega$ coupling'' terms in the separated Teukolsky equations 
and has been discussed in~\cite{PT1973,gr-qc/0610122}. However,  as emphasized 
in~\cite{gr-qc/0610122}, this coupling has only a small effect on the 
$(\l,m) = (2,2)$ waveform. We shall neglect it and consider only the 
(positive-frequency) QNM modes having the same values of $(\l,m)$ as 
the considered multipolar waveform $h_{\l m}$ (i.e. $(2,2)$ in the present paper).

On the other hand, contrary to other recent implementations of the EOB 
approach~\cite{gr-qc/0610122,arXiv:0704.1964,arXiv:0706.3732}, 
we shall use a matching comb with {\it five} teeth ($p=2$) and 
{\it five} (positive-frequency) QNMs 
$\sigma_{\l mn}^{+}= \alpha_{\l mn}^{+} + {\i}\omega_{\l mn}^{+}$, with 
$\l =2$, $m=+2$, and $n=0,1,2,3,4$. To estimate the values (as functions of the
mass and spin of the final black hole) of the damping time and the oscillation 
frequency of each mode we did the following:
(i) for the first three modes we used the approximate fitting formulas given in
Appendix~E of Ref.~\cite{gr-qc/0512160}; while, (ii) for the fourth and fifth
modes (i.e. $n=3,4$) we noticed that the graphic results
of~\cite{Onozawa:1996ux} (notably his most relevant Fig.~4) exhibit an 
{\it approximate linearity} of the complex QNM frequency $\sigma_{\l mn}^{\pm}$ 
as a function of the overtone number $n$. [Indeed, the corresponding points 
in the complex $\sigma$ plane are approximately aligned.] We  then exploited 
this approximate linearity to express the needed $n=3$ and $n=4$ complex 
frequencies as linear combinations of the above-discussed $n=1$ and $n=2$ ones.

\item Concerning the {\it matching}, on a multi-toothed ``comb'', of the inspiral-plus-plunge 
waveform to the ring-down one we need to specify the two parameters
defining such a comb, namely the central``matching'' time $t_m$,
and the spacing between the teeth of the comb: $\delta = \Delta t/4$\footnote{
Note that in~\cite{arXiv:0705.2519} we used the letter $\delta$ to denote
the full width $\Delta t$ of the comb.}. In conformity with the basic idea
proposed in the original EOB paper~\cite{Buonanno:2000ef} we choose as
central matching time $t_m$ the so-called ``EOB light-ring crossing'' time;
i.e., the EOB dynamical time when the EOB orbital frequency $\Omega$ reaches
its maximum. See~\cite{arXiv:0705.2519} for a detailed  discussion
of why such a choice is physically preferred. Concerning the choice of the
comb spacing $\delta$, we expect from~\cite{arXiv:0705.2519} that a value 
of order $\delta = (7.2 M)/4 = 1.8 M$ will be good. Below, we shall explore
values near this one.

\end{itemize}

\section{Comparing the NR waveform to EOB ones}
\label{sec:comparing}

As explained in Sec.~\ref{sec:nr}, the basic NR 
data that we shall consider is a time-series giving the quadrupolar 
[$(\l,m)=(2,2)$, Zerilli-Moncrief-normalized] metric waveform 
$\Psi_{22}^{\rm NR}$ as a function of the NR time 
variable\footnote{As mentioned in Sec.~\ref{sec:nr}, we use the waveform 
extracted at a (coordinate) radius $R_{\NR} = 120M_c\simeq 120M$, and 
$t_{\rm NR}$ is the time of the ``observer'' located at the latter radius.} 
$t_{\rm NR}$ (measured in units of $M\equiv m_1 + m_2$). 
$\Psi_{22}^{\rm NR}(t_{\rm NR})$ is a complex number. The NR results 
consist of the real and imaginary parts of $\Psi_{22}^{\rm NR}$.
It is, however, more convenient to decompose the complex waveform 
in modulus (or amplitude) and phase, say

\be
\Psi_{22}^{\rm NR}(t_{\rm NR}) = A_{22}^{\rm NR}(t_{\rm NR}) 
\exp \left( - \i \phi_{22}^{\rm NR}\left(t_{\rm NR}\right)\right) .
\ee

The $2 \pi$ ambiguity in the phase is fixed by starting with the principal
value of the argument of $\Psi_{22}^{\rm NR}$ at the beginning of the NR simulation,
and then keeping track of the $2 \pi$ turns as the waveform continuously
unfolds. 

One can then compute the gravitational wave (GW) frequency as a function of
time by (numerically) differentiating the GW phase

\be
\omega_{22}^{\rm NR}(t_{\rm NR}) = \frac{d \phi_{22}^{\rm NR}}{d t_{\rm NR}} .
\ee
[It can equivalently be obtained by computing the imaginary part of the
logarithmic time derivative of $\Psi_{22}^{\rm NR}(t_{\rm NR})$.]

As emphasized in~\cite{arXiv:0711.2628}, another useful diagnostics of GW radiation
is the GW phase acceleration \hbox{$\alpha = d \omega/ dt=d^2\phi/dt^2$} considered as a function of
the GW frequency $\omega$. However, because of the presence of some 
additional high-frequency wiggles in $\phi$ and $\omega$ in the NR data, we
shall not consider here the phase-acceleration curve $\alpha(\omega)$.
Instead, we shall directly compare the numerical GW amplitude, phase and frequency
to their analytical, EOB counterparts.

The integration of the basic EOB dynamical equations (written in~\cite{arXiv:0711.2628})
gives, for each chosen value of the EOB ``flexibility parameters'' (notably $a_5$
and $v_{\rm pole}$), several important time series, and notably:
(i) the EOB orbital frequency $\Omega(t_{\rm EOB})$, where $t_{\rm EOB}$ is the
EOB dynamical time scale (measured in units of $M$); (ii) the new, resummed 
 {\it matched} $3^{+2}$-PN-accurate quadrupolar EOB waveform 
 $\Psi_{22}^{\rm EOB}(t_{\rm EOB})$; then, from
 the latter, one can define (as for the NR case) the corresponding EOB amplitude,
 $A_{22}^{\rm EOB}(t_{\rm EOB})$, EOB phase, $ \phi_{22}^{\rm EOB}(t_{\rm EOB})$,
 and EOB frequency $\omega_{22}^{\EOB}(t_{\EOB})$.
To compare the NR and EOB phase time-series 
$ \phi_{22}^{\rm NR}(t_{\rm NR})$ and $\phi_{22}^{\rm EOB}(t_{\rm EOB})$
one needs to shift, by additive constants, both one of the time variables, and 
one of the phases. In other words, we need to determine $\tau$ and $\alpha$ such that
the ``shifted'' EOB quantities 

\be
t'_{\rm EOB}=t_{\rm EOB} + \tau \ , \quad
\phi_{22}^{'\rm EOB} = \phi_{22}^{\rm EOB} + \alpha
\ee
``best fit'' the NR ones. One convenient way to do so is first to ``pinch''
the EOB/NR phase difference at two different instants (corresponding to two
different frequencies). More precisely, one can choose 
two NR times  $t_1^{\rm NR},t_2^{\rm NR}$, which determine  two corresponding
 GW frequencies\footnote{
Alternatively, one can start by giving oneself $\omega_1,\omega_2$ and determine the NR
instants $t_1^{\rm NR},t_2^{\rm NR}$ at which they are reached.}
$\omega_1= \omega_{22}^{\rm NR}(t_1^{\rm NR})$, $\omega_2=\omega_{22}^{\rm NR}(t_2^{\rm NR})$,
and then find the time shift $\tau(\omega_1,\omega_2)$ such that the shifted EOB
phase difference, between $\omega_1$ and $\omega_2$,
$\Delta\phi^{\rm EOB}(\tau) \equiv 
\phi_{22}^{'\rm EOB}(t_2^{'\rm EOB}) - \phi_{22}^{'\rm EOB}(t_1^{'\rm EOB})=
\phi_{22}^{\rm EOB}(t_2^{\rm EOB}+ \tau) - \phi_{22}^{\rm EOB}(t_1^{\rm EOB}+\tau)$
is equal to the corresponding (unshifted) NR phase difference
$\Delta\phi^{\rm NR} \equiv \phi_{22}^{\rm NR}(t_2^{\rm NR}) - \phi_{22}^{\rm NR}(t_1^{\rm NR})$. This yields one equation for one unknown ($\tau$), and (uniquely)
determines a value $\tau(\omega_1,\omega_2)$ of $\tau$. [Note that the 
$\omega_2 \to \omega_1 =\omega_m$ limit of this procedure yields the 
one-frequency matching procedure used in~\cite{arXiv:0710.0158}.] After having
so determined $\tau$, one can uniquely define a corresponding best-fit phase shift 
$\alpha(\omega_1,\omega_2)$ by
requiring that, say, 
$\phi_{22}^{'\rm EOB}(t_1^{'\rm EOB}) \equiv \phi_{22}^{\rm EOB}(t_1^{'\rm EOB})
+ \alpha = \phi_{22}^{\rm NR}(t_1^{\rm NR})$.

Having so related the EOB time and phase variables to the NR ones we can 
straigthforwardly compare all the EOB time series to their NR correspondants.
In particular, we can compute the (shifted) EOB--NR phase difference
\be
\label{deltaphi}
\Delta^{\omega_1,\omega_2}\phi_{22}^{\rm EOB NR} (t_{\rm NR}) \equiv \phi_{22}^{'\rm EOB}(t'^{\rm EOB}) - \phi_{22}^{\rm NR}(t^{\rm NR}).
\ee
In the following we will chose two matching
instants (and corresponding frequencies) that take place during 
late inspiral and  plunge,
namely:  $t_1^{\rm NR}=999.72$,  $t_2^{\rm NR}=1494.94$ corresponding to 
$\omega_1=0.06815$, $\omega_2=0.2457$ (all expressed in $M$ units).

To numerically implement the EOB/NR comparison we need to choose some
values for the various ``flexibility parameters'' of the EOB framework.
We have summarized above what are these parameters, and we have already indicated
the values we chose for some of them. Among the remaining ones that need to be chosen,
the two most crucial ones are $a_5$ and $v_{\rm pole}$. 
Recently, Damour and Nagar have shown,
by using some of the data published in~\cite{arXiv:0710.0158}, that the
{\it inspiral waveform} (for GW frequencies smaller than about $0.14/M$) could
be remarkably well matched by the EOB one if one chose values of $a_5$ and $v_{\rm pole}$
following the rather precise correlation plotted in the upper panel of Fig.~3
in~\cite{arXiv:0711.2628}. 
%
%
\begin{figure*}[t]
  \begin{center}
    \includegraphics[width=85 mm]{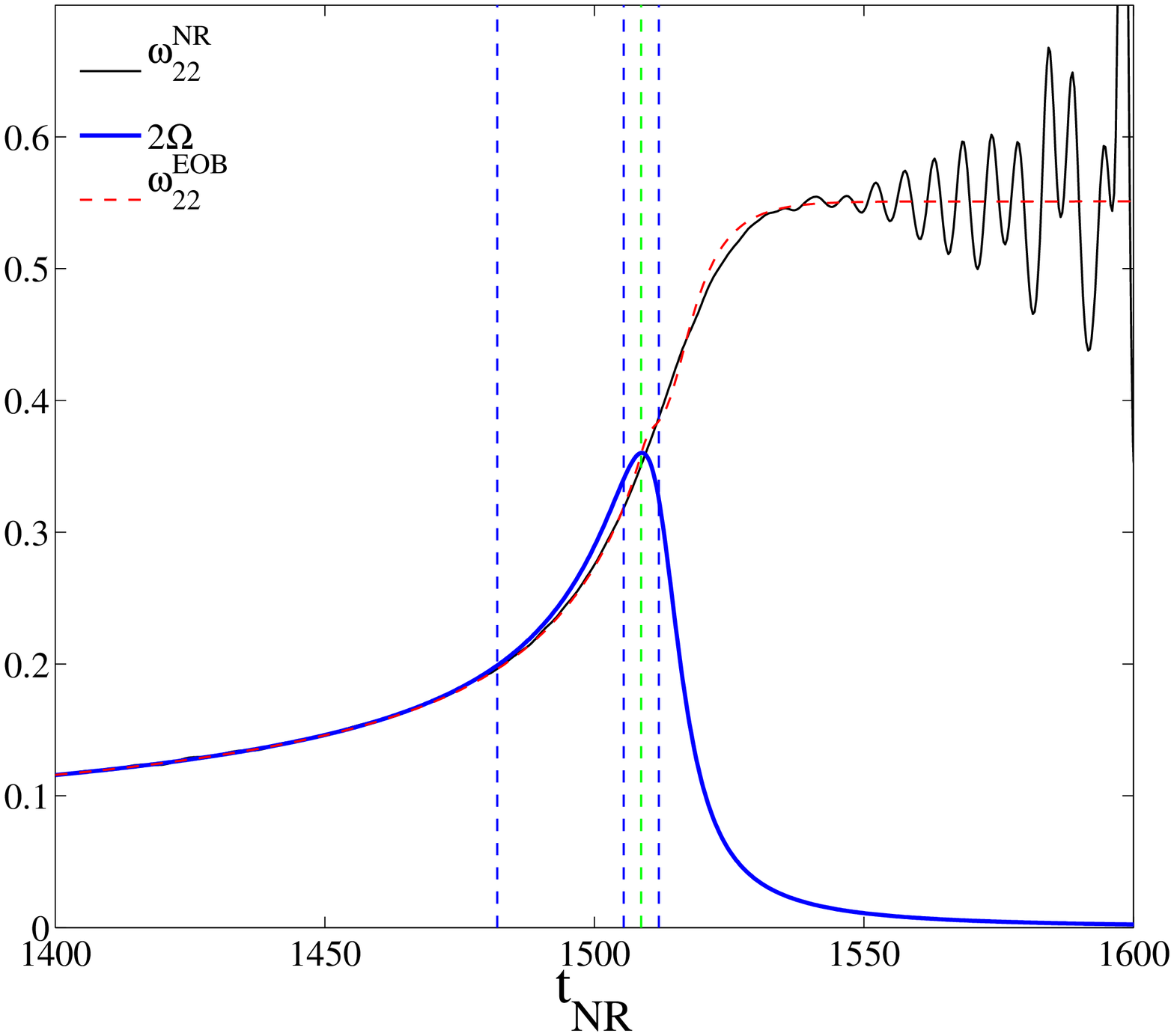} 
    \includegraphics[width=85 mm]{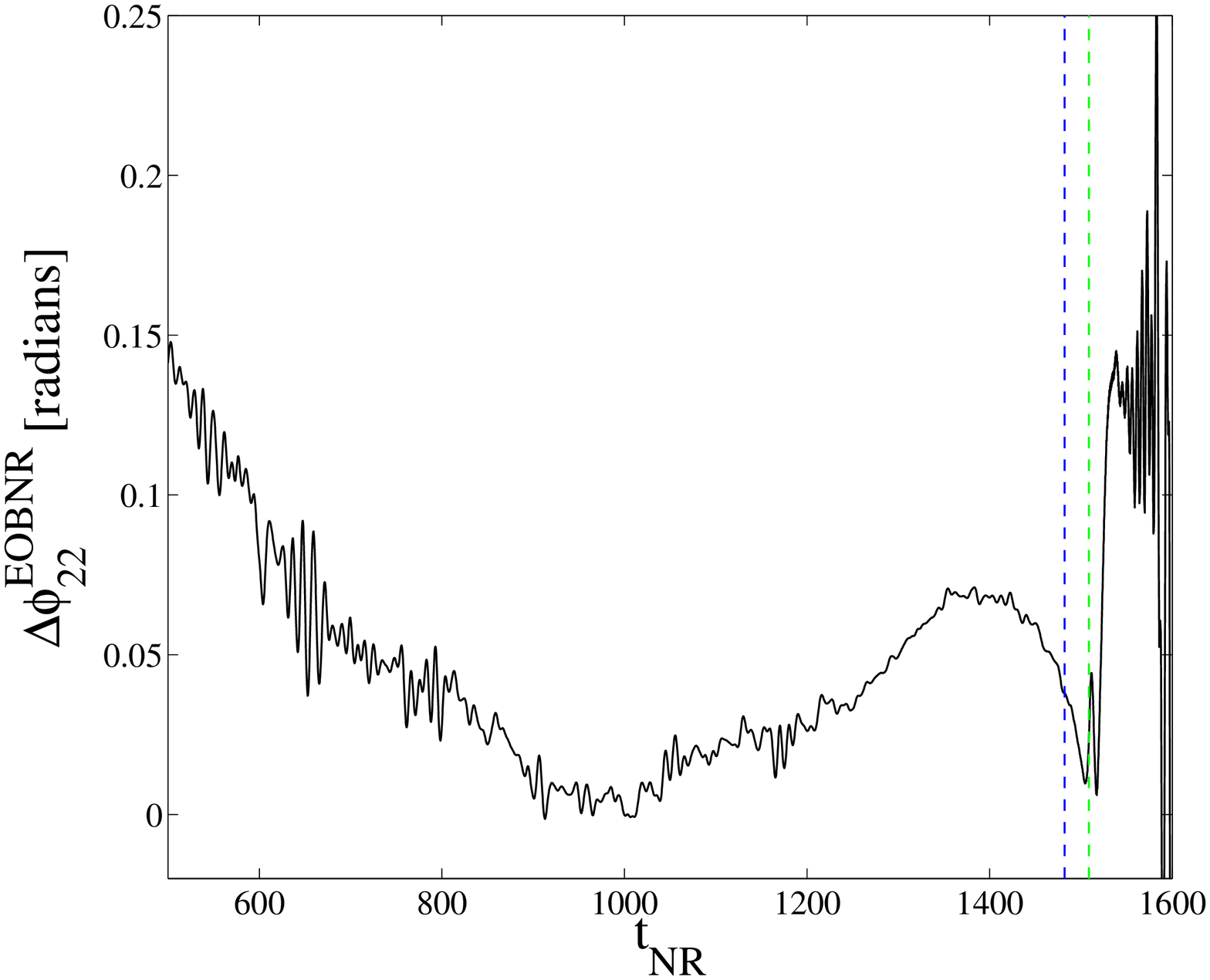} \\ 
    \includegraphics[width=85 mm]{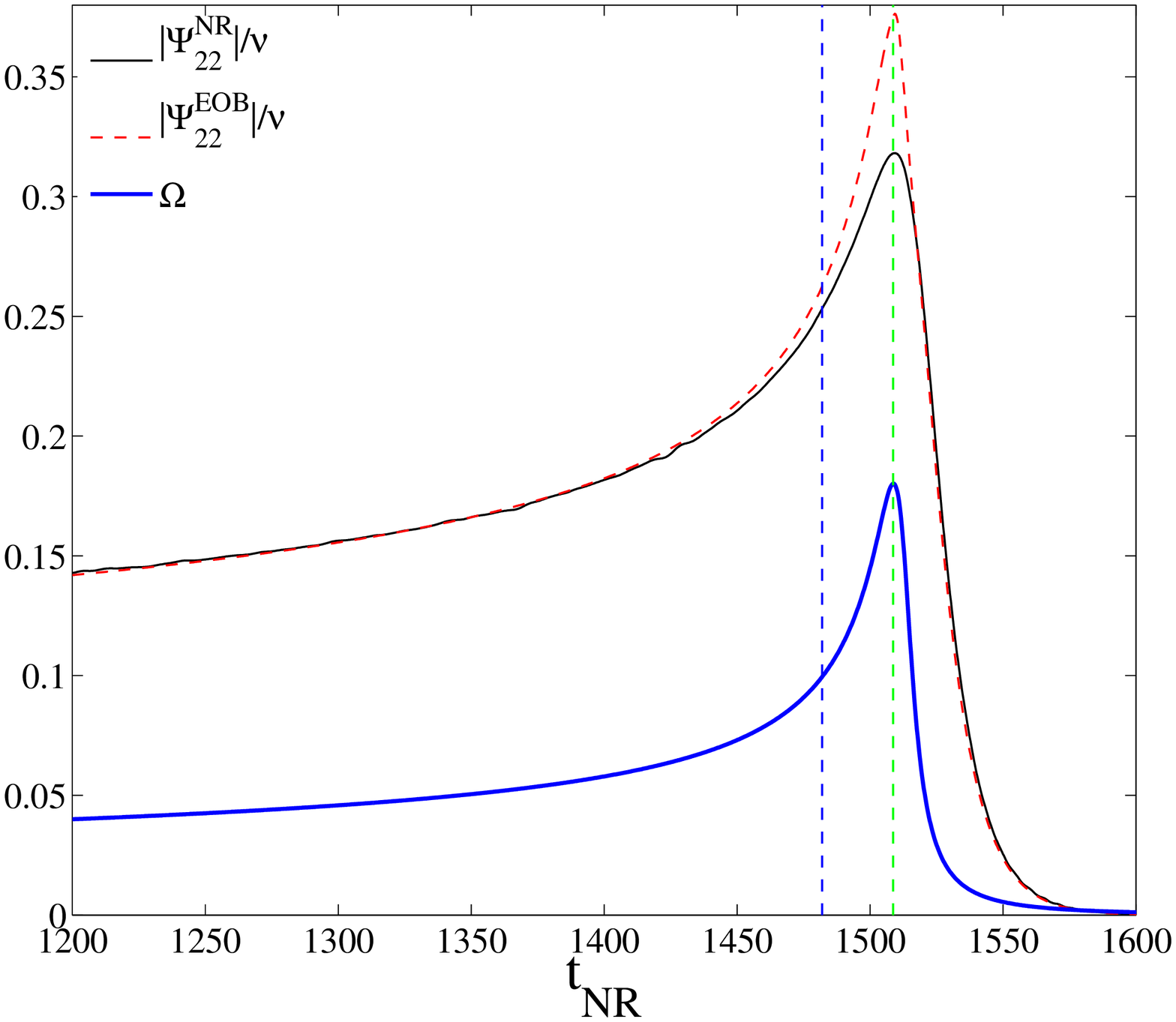} 
    \includegraphics[width=85 mm]{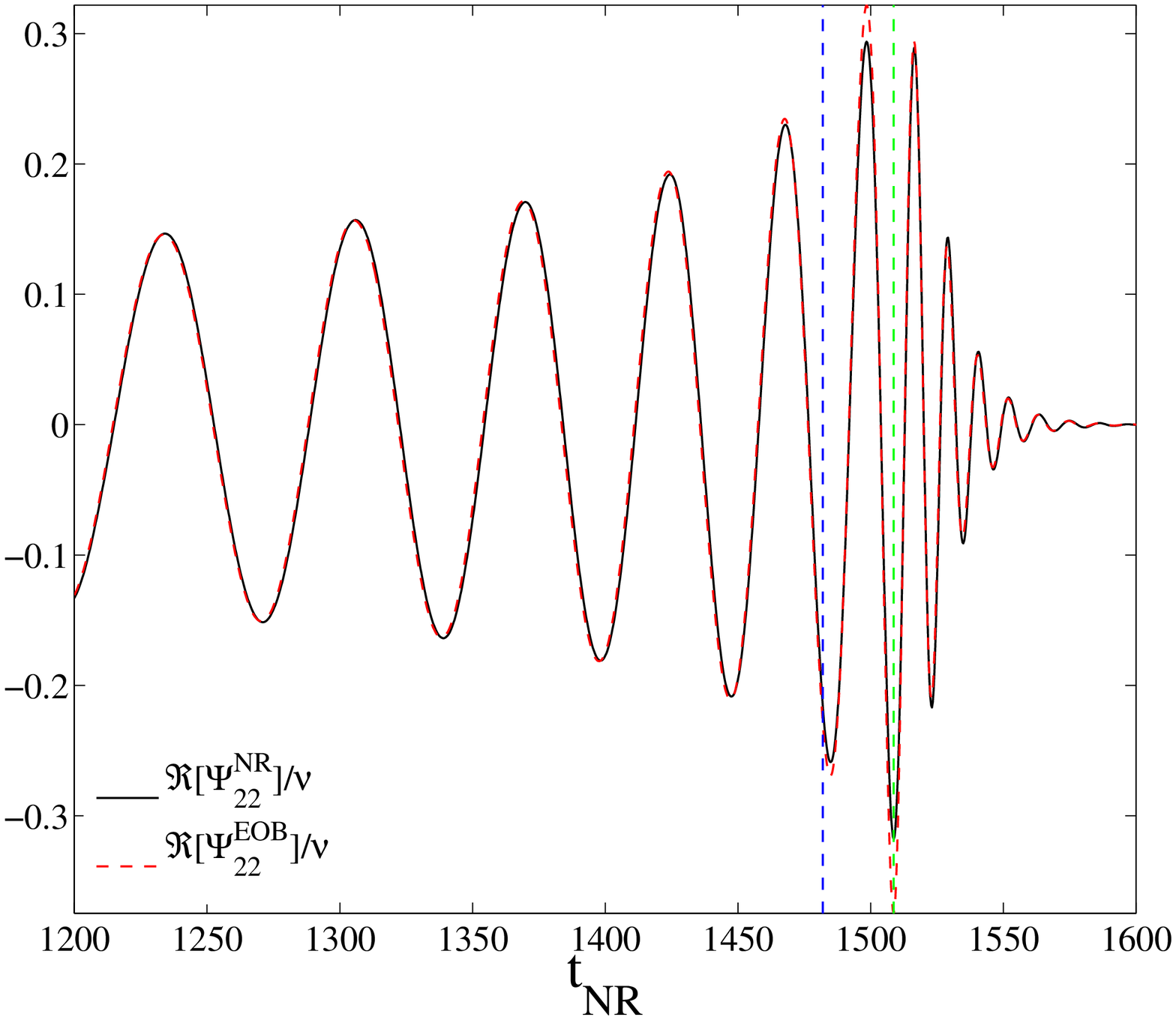} 
  \end{center}
  \vspace{-4mm}
  \caption{\label{label:fig1}Comparison between EOB and NR waveforms for
    $a_5=25$ and $v_{\pole}=0.6241$: frequencies (top--left), phase difference
    (top--right), amplitudes (bottom--left) and real parts (bottom--right) 
    of the two gravitational waveforms. The vertical line at $t_{\NR}=1509$
    locates the maximum of (twice) the orbital frequency $\Omega$ 
    (alias the ``EOB-light-ring'') and indicates the center of our matching 
    comb (whose total width is indicated by the two neighboring vertical 
    lines in the top--left panel). The vertical dashed line at $t^{\NR}=1482$ 
    indicates the crossing time of the adiabatic LSO orbital frequency 
    ($\Omega_{\rm LSO}= 0.1003$).}
\end{figure*}
Here, as we are exploring a different physical regime
(late inspiral, plunge and coalescence, with GW frequencies mostly larger
than about $0.1/M$), and comparing to a different set of numerical data, we shall not
a priori impose the precise correlation between  $a_5$ and $v_{\rm pole}$
found in~\cite{arXiv:0711.2628}. However, we shall make use of some previous
results suggesting a preferred range for the values of $a_5$. On the one hand,
Ref.~\cite{arXiv:0706.3732} showed that the {\it faithfulness} (in the sense 
of Sec.~VIA of~\cite{gr-qc/0211041}) of (restricted) EOB waveforms against 
NASA-Goddard NR coalescence waveforms was largest when $a_5$ belonged to 
some rather wide interval (which also depended on the considered mass ratio). 
See Fig.~2 (right panel)
in~\cite{arXiv:0706.3732} from which one might conclude that $a_5$ lies
probably between $\sim 10$ and $\sim 100$. Buonanno et al. then chose
$a_5 = 60$ as ``best fit'' value. On the other hand, Ref.~\cite{arXiv:0711.2628}
found that the phase agreement between (resummed) EOB waveforms and a rather long inspiral
NR waveform was at its best when $a_5$ lied in a similarly wide interval
(between $\sim 10$ and $\sim 80$) centered around $a_5 \sim 40$. In view of these
results we shall focus, in the following, on two representative values of $a_5$,
namely $a_5 = 25$ (representative of the leftward-side of preferred $a_5$ values),
and $a_5 = 60$ (representative of the rightward-side of preferred $a_5$ values, and
chosen as best value by~\cite{arXiv:0706.3732}). We have also checked that the
values of $a_5$ between 25 and 60 lead (with appropriate choice of $v_{\rm pole}$)
to results that are at least as good as the ones we shall exhibit below.

\subsection{Comparing NR to resummed EOB for $a_5 = 25$}

At this stage we have essentially fixed all the flexibility of the
EOB formalism apart from the choices of $v_{\rm pole}$, and of the 
comb spacing $\delta$. Among these two parameters, only the former
one, $v_{\rm pole}$, is important for getting a very accurate phase
agreement between EOB and NR. When $a_5 = 25$, we found (by trial and error)
that\footnote{Though we did not investigate thoroughly what ``error bar'' 
can be put on such a ``best'' value of $v_{\pole}$, the numerical studies we did
indicate that a change of $\pm 2$ on the last (i.e. fourth) digit that we
quote is sufficient to entail a visible worsening of the phase difference
$\Delta\phi_{22}^{\EOB\NR}$.} $v_{\rm pole}=0.6241$ 
(together with $\delta = 1.7 M_{\rm f}$ which is, however, less crucial) 
yields an excellent EOB/NR agreement. 
We exhibit our results in the four panels of Fig.~\ref{label:fig1}.

The top-left panel of Fig.~\ref{label:fig1} compares the NR GW 
frequency both to the (matched) EOB GW frequency, and to twice 
the orbital frequency. The time axis is 
$t_{\NR}$, and/or  (see above) $t'_{\EOB}=t_{\EOB} + \tau$ 
(with $\tau=-2032M$ for the present case).
The vertical lines on the right indicate the center and the 
outlying ``teeth'' of our matching comb, which is,
as explained above, centered on the maximum of the EOB orbital 
frequency (also called ``EOB-light-ring''). The interval between 
the two vertical lines (LSO and ``EOB-light-ring'') defines the ``plunge''.
The dashed vertical line on the left (at $t_{\NR}= 1482$) indicates 
the crossing time of the adiabatic Last-Stable-Orbit 
($\omega$-LSO in the sense of~\cite{Buonanno:2000ef}).
Note that the three frequencies are initially close to each other, but that,
later, $2\Omega$ separates from $\omega_{22}^{\NR}$ and $\omega_{22}^{\EOB}$,
which continue to be in very close agreement, except for a slight discrepancy
around merger, which, within the EOB approach, is conventionally supposed to 
take place at the maximum of $\Omega$. 
Note also the good agreement between the EOB GW frequency
during the ringdown plateau, and the average of the NR one. 
As discussed in Sec.~\ref{sec:nr} the values for the mass 
and dimensionless spin of the final black hole
that we used (together with \cite{gr-qc/0512160})  
to compute the QNMs frequencies are: $M_{\rm f}^{\rm ring}=0.959165 M$,
$j_{\rm f}^{\rm ring}= 0.684639$.

The top-right panel of Fig.~\ref{label:fig1} shows the EOB-NR phase 
difference, Eq.~(\ref{deltaphi}), (``pinched'' at the two instants, 
$t_1^{\NR}$, $t_2^{\NR}$, given above). 
It is remarkable that the (two-sided~\footnote{As the reference level of
any phase difference $\Delta\phi$ is arbitrary, it is convenient to use
a ``middle'' reference level such that $\Delta\phi(t)$ varies between 
$-\varepsilon$ and $+\varepsilon$ over the considered interval. We
refer to $\pm\varepsilon = \pm1/2[\max(\Delta\phi)-\min(\Delta\phi)]$
as the {\it two-sided} phase difference.}) EOB-NR phase difference over the time
interval $(639 M, 1524 M)$ (which covers about 12~GW cycles of
inspiral, plunge, and early ring-down) is smaller than about 
$ \pm \frac{1}{2} 0.068$ radians, which corresponds to $\pm 0.005$ GW cycles.

The bottom-left panel of Fig.~\ref{label:fig1} compares the NR GW 
amplitude to the resummed $3^{+2}$-PN accurate EOB one. It also 
shows the orbital frequency $\Omega$ as an aid to locate the merger.
 One notices a very good agreement between the two amplitudes. 
During the interval $(1100M,1400M)$ the fractional EOB-NR amplitude
difference varies between $-1\%$ and $+1\%$.
After $t_{\NR}=1400M$, this fractional difference increases 
from $+1\%$ to a maximum of $+18\%$ (reached at $t_{\NR}\simeq 1509M$) 
and then decreases to take values of order $-5\%$ during the 
observationally relevant part of the ringdown.
Note also that the NR equal-mass amplitude (divided by $\nu$, i.e. by $\mu$)
time series is qualitatively, and even quantitatively, very similar to
the corresponding NR test-mass amplitude time series shown in Fig.~3
of~\cite{arXiv:0705.2519}. For instance, the value of the maximum 
amplitude is $\sim 0.3$ in both cases. A similar qualitative, 
but {\it not} quantitative, parallelism exists for the two 
corresponding frequency time series (the $\nu=1/4$ frequency
levelling off at a higher ``plateau'').

%
%
\begin{figure*}[t]
  \begin{center}
    \includegraphics[width=85 mm]{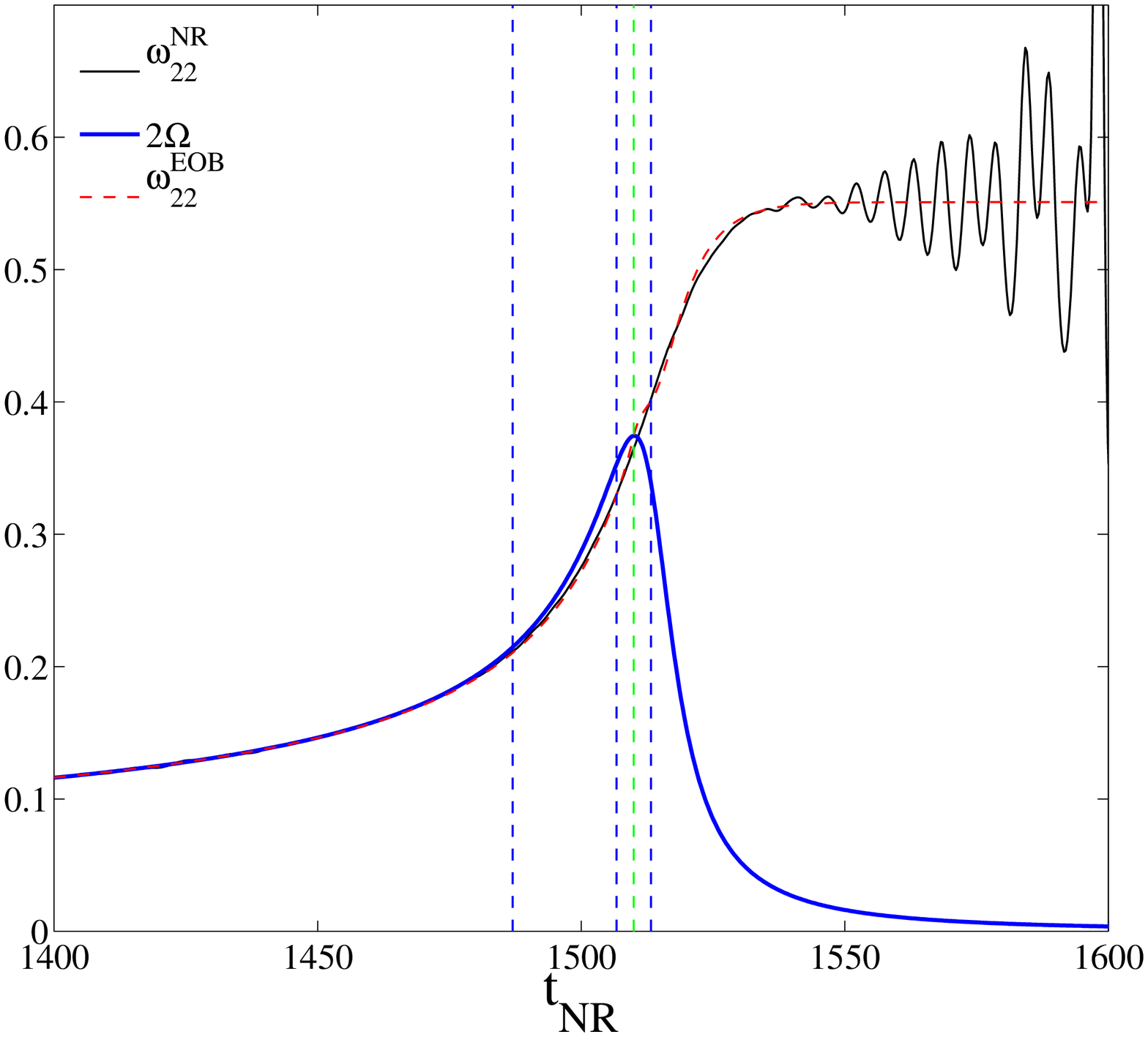} 
    \includegraphics[width=85 mm]{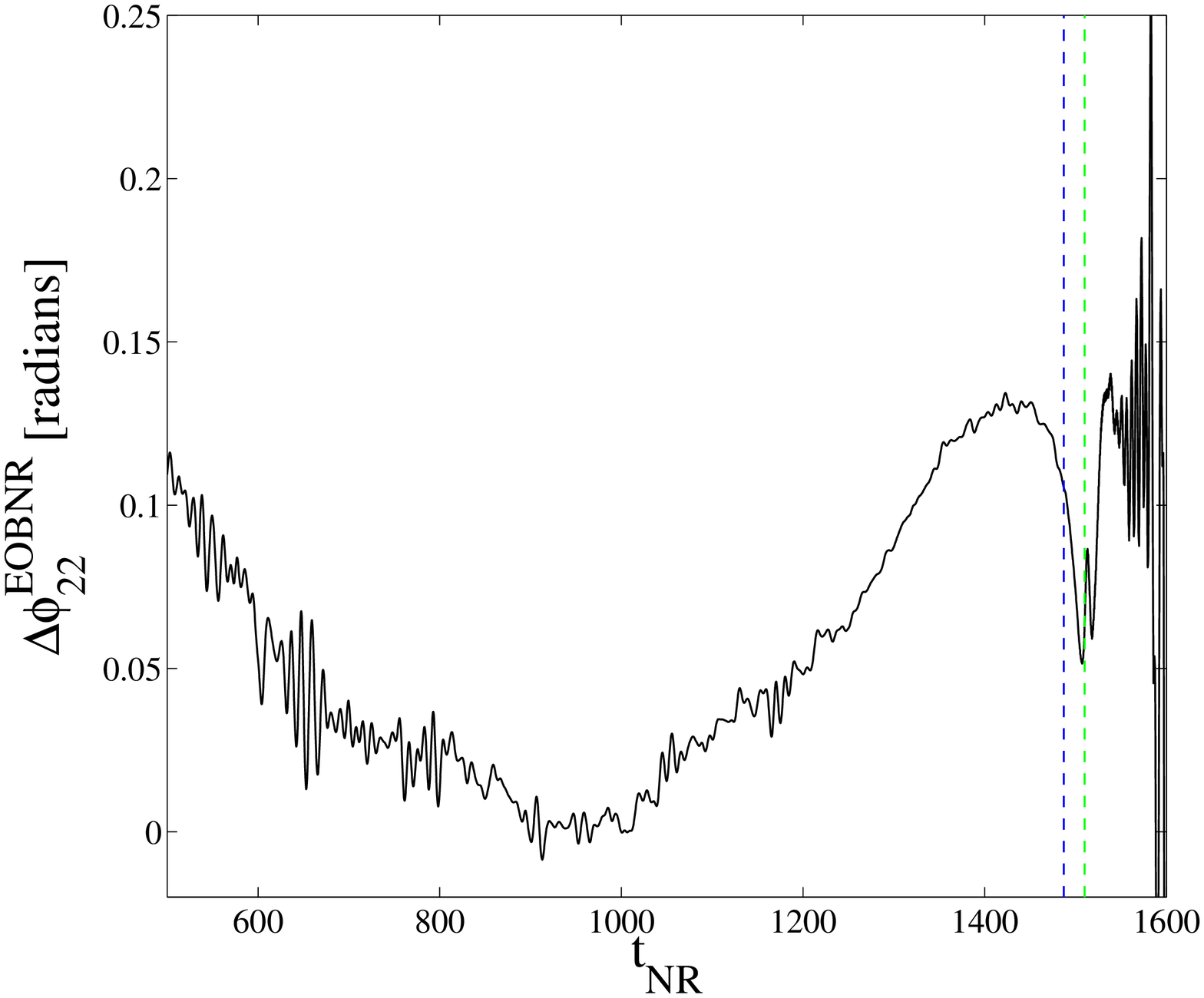} \\ 
    \includegraphics[width=85 mm]{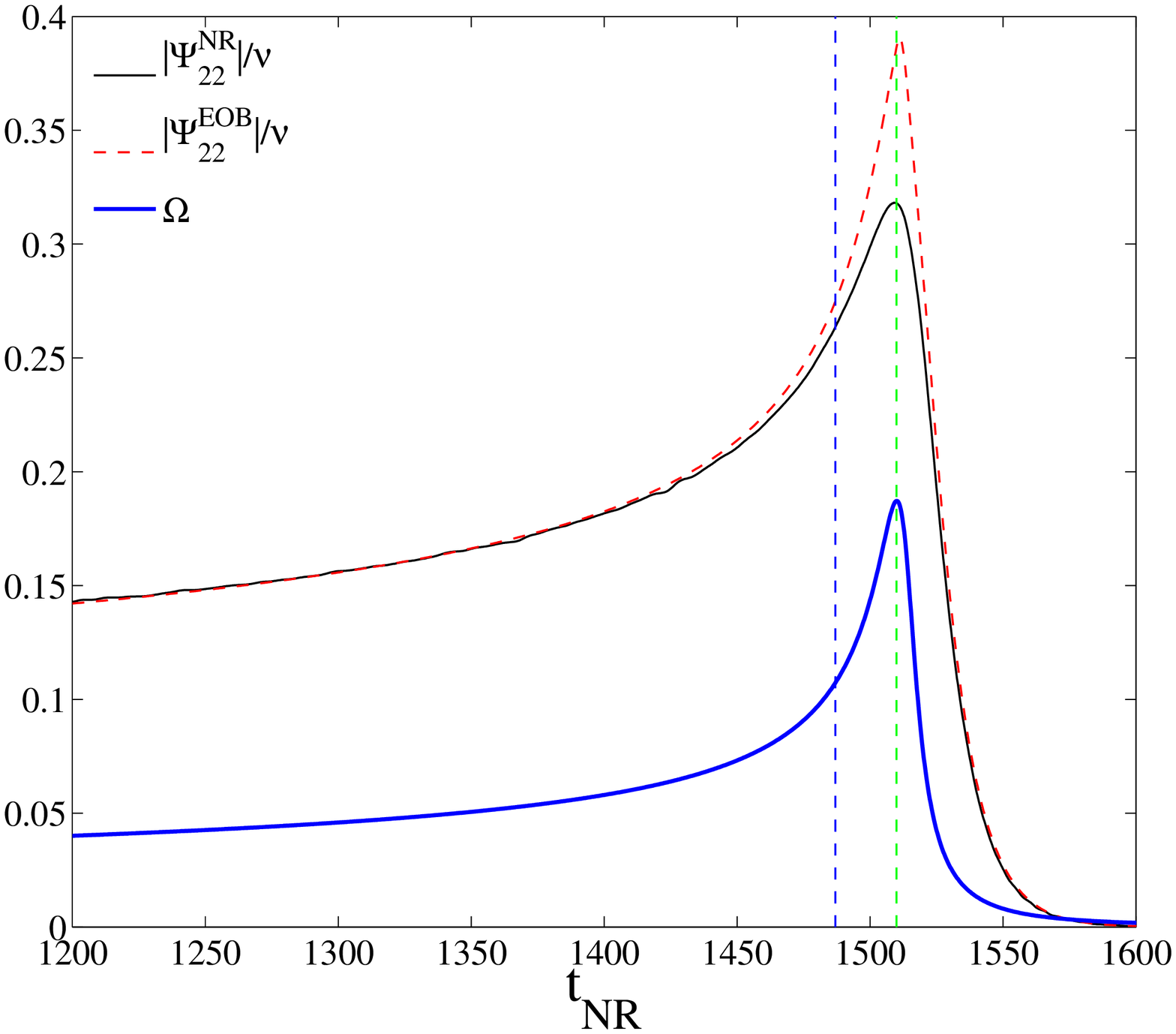} 
    \includegraphics[width=85 mm]{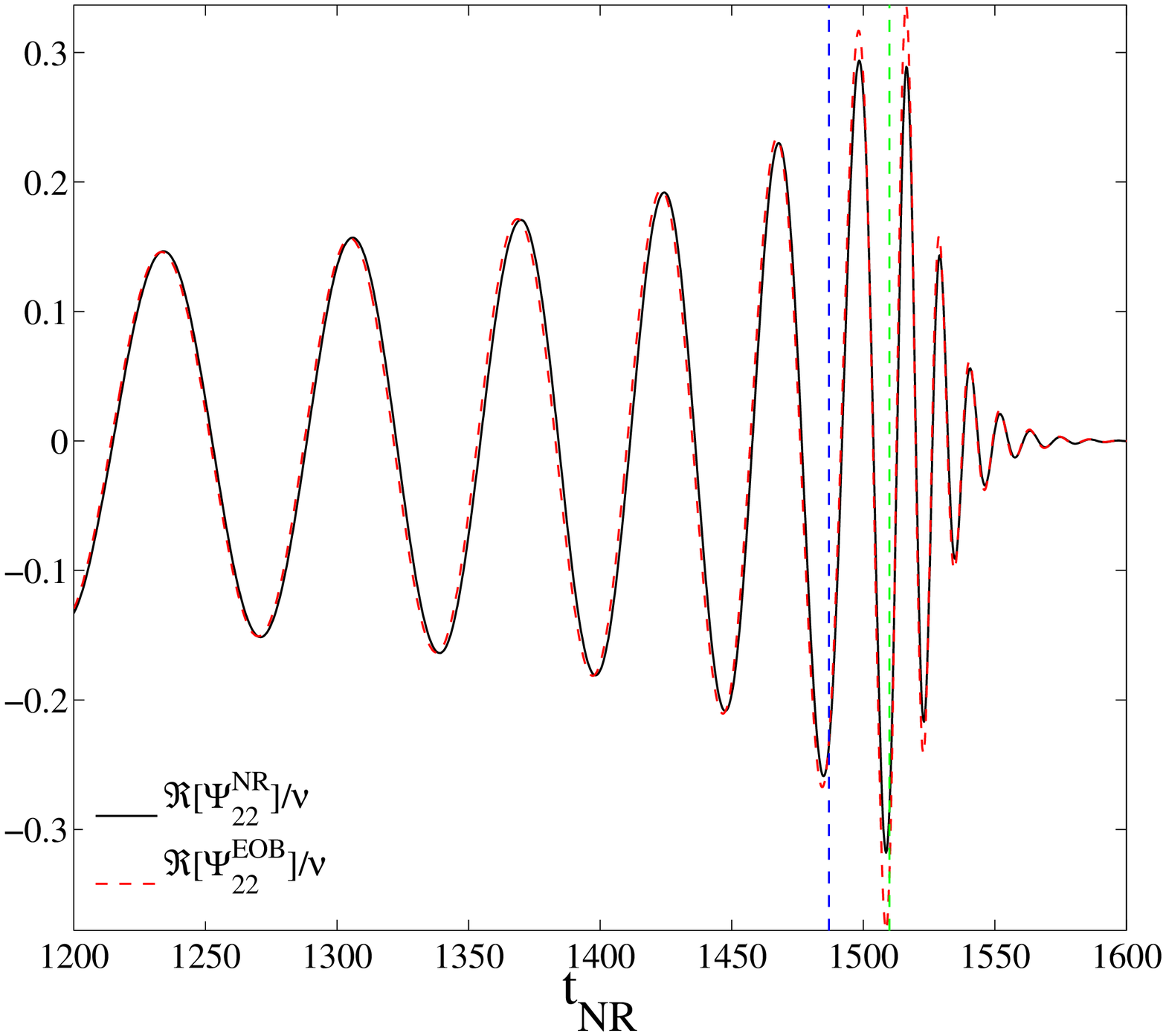} 
  \end{center}
  \vspace{-4mm}
  \caption{\label{label:fig2}Comparison between EOB and NR waveforms 
   for $a_5=60$, $v_{\rm pole}=0.5356$: frequencies (top--left), phase difference
   (top--right), amplitudes (bottom--left) and real parts (bottom--right) 
   of the two gravitational waveforms. The vertical line at $t_{\NR}=1510$
   locates the maximum of (twice) the orbital frequency $\Omega$ 
   (alias the ``EOB-light-ring'') and indicates the center of our matching 
   comb (whose total width is indicated by the two neighboring vertical 
   lines in the top--left panel).
   The vertical dashed line at $t^{\NR}=1487$ indicates the crossing 
   time of the adiabatic LSO orbital frequency ($\Omega_{\rm LSO}= 0.1081$).}
\end{figure*}

Finally, the bottom-right panel of Fig.~\ref{label:fig1} compares 
the real parts of the NR and EOB waveforms. 
The two vertical lines delimit the interval between LSO and ``EOB-light-ring''.
Again the agreement 
between the two waveforms is impressive. Note that this last panel shows only 
the late inspiral, plunge and ringdown. From the panel showing 
the phase difference, one can gather that the agreement stays 
as impressive over a much longer time span  of order $1000M$ 
(essentially from $t_{\NR} \sim 500 M$ to the end of ringdown).

\subsection{Comparing NR to  resummed EOB  for $a_5 = 60$}

%
%
\begin{figure*}
  \begin{center}
    \includegraphics[width=85 mm]{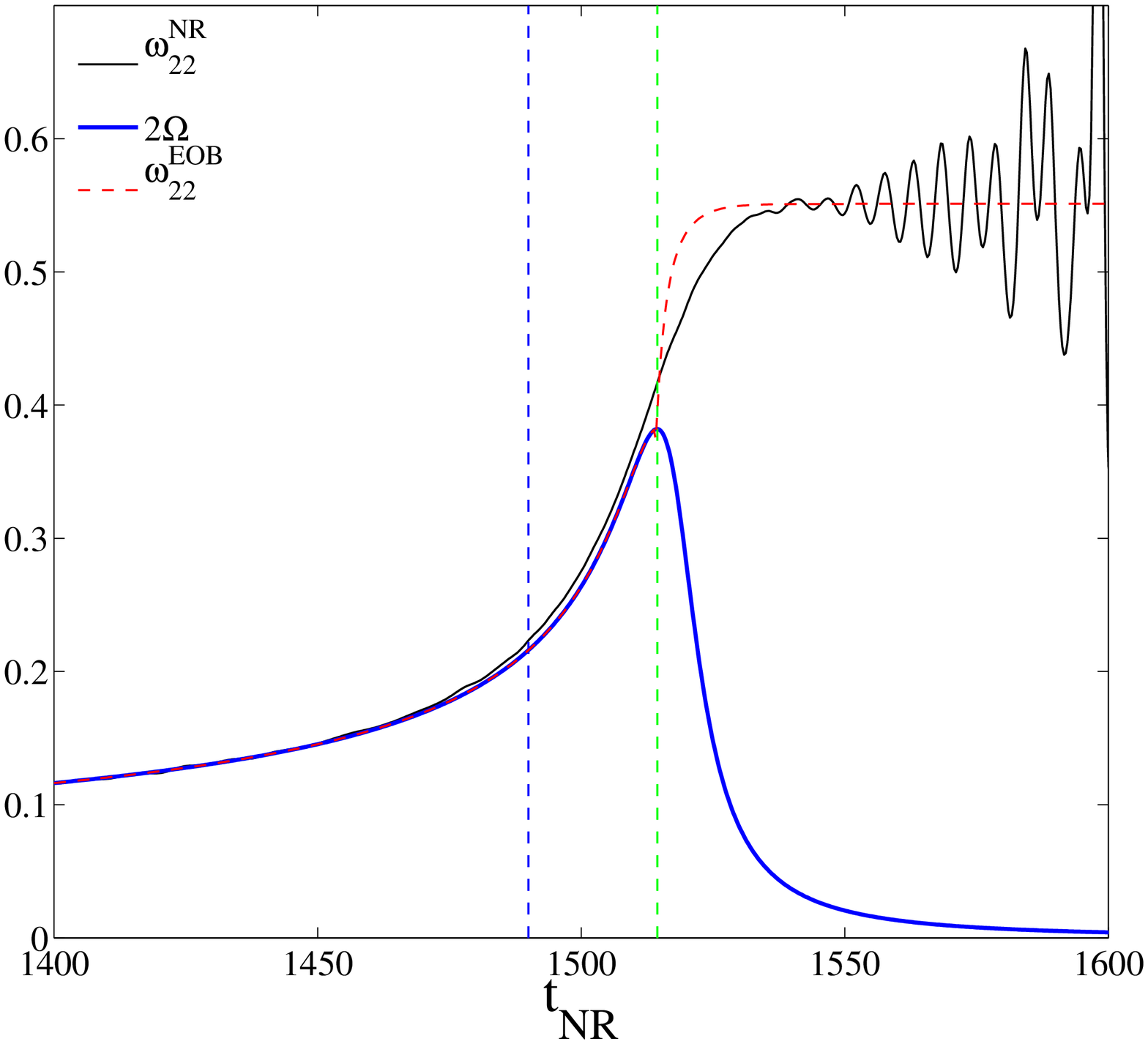} 
    \includegraphics[width=85 mm]{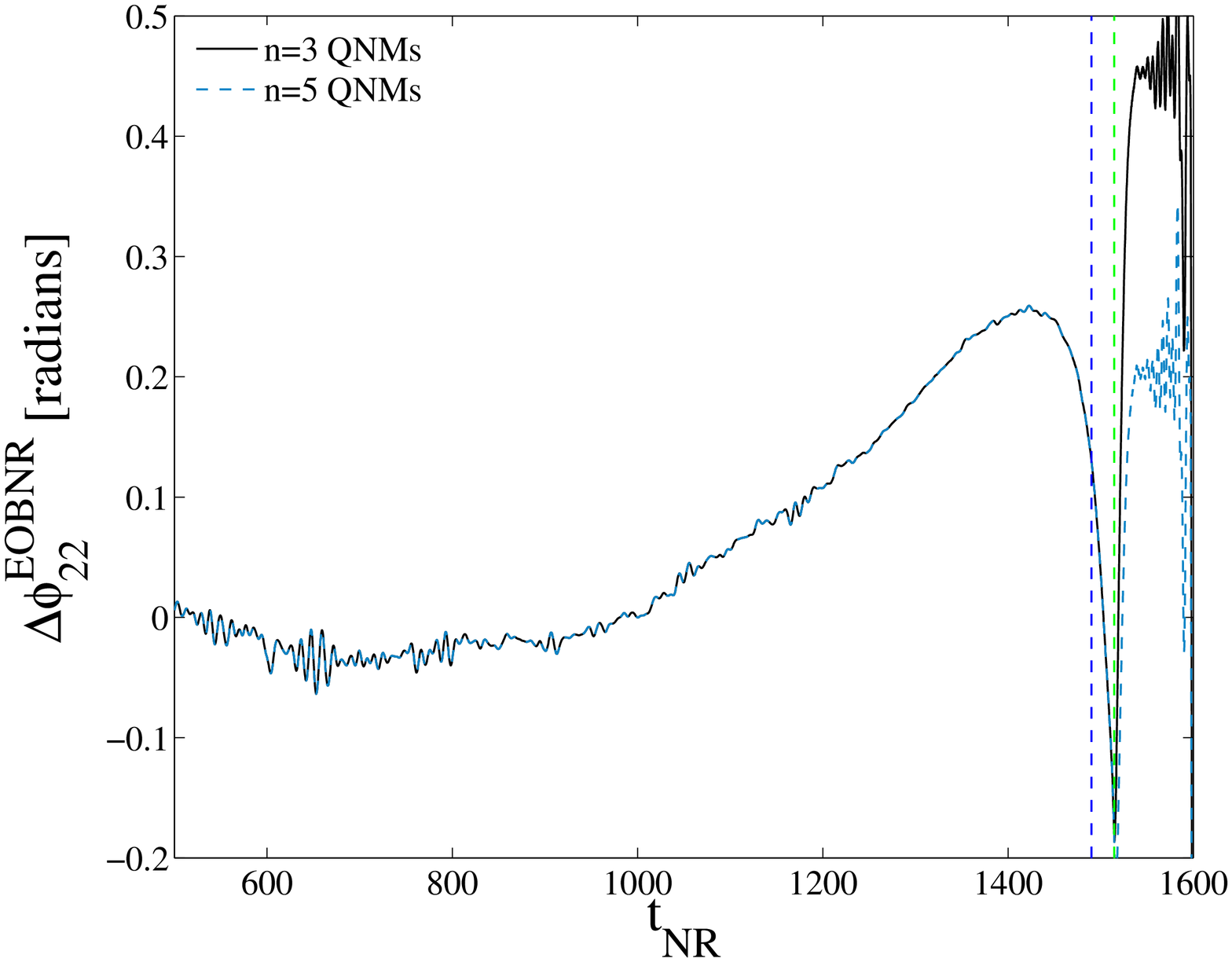} \\ 
    \includegraphics[width=85 mm]{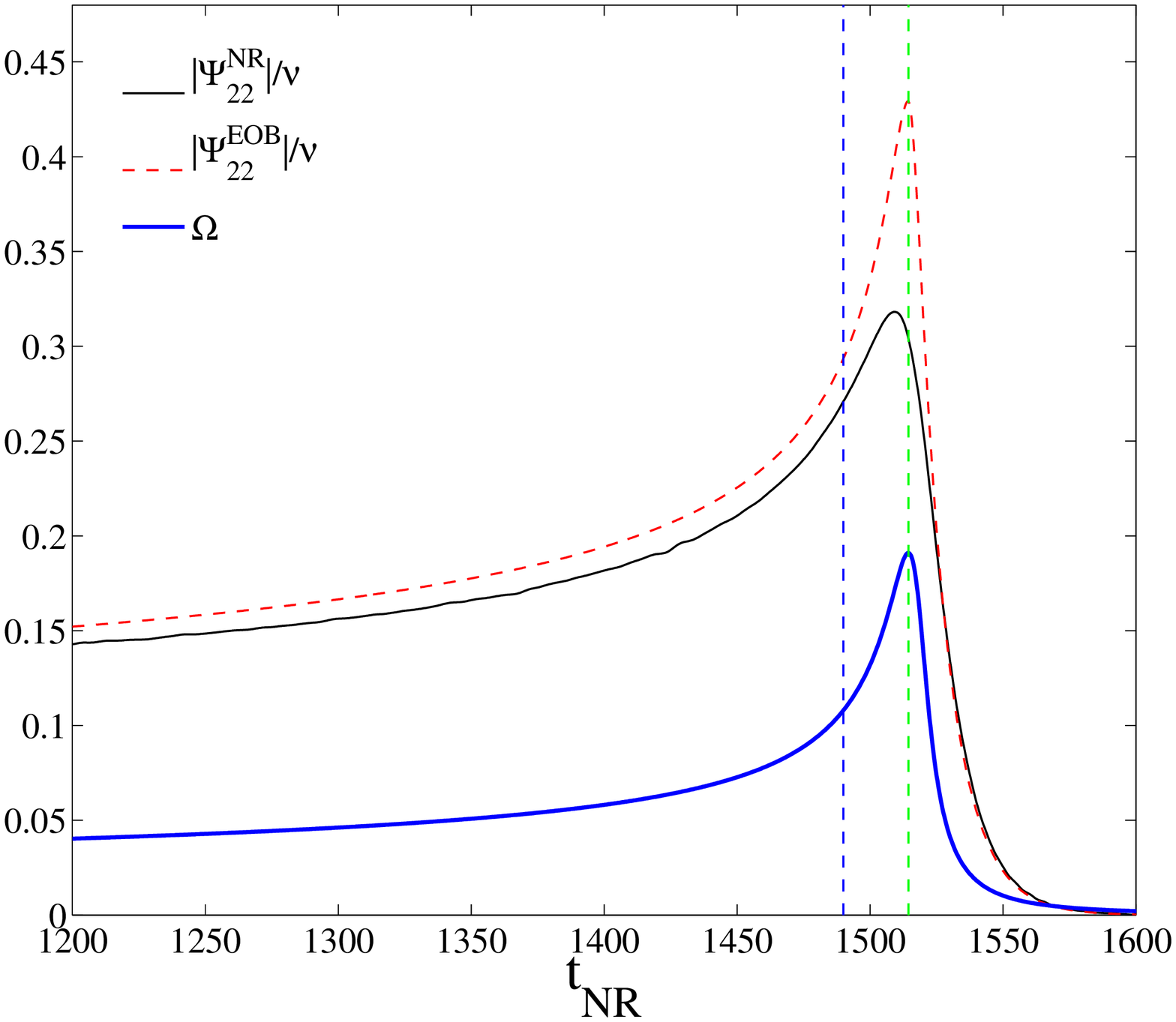} 
    \includegraphics[width=85 mm]{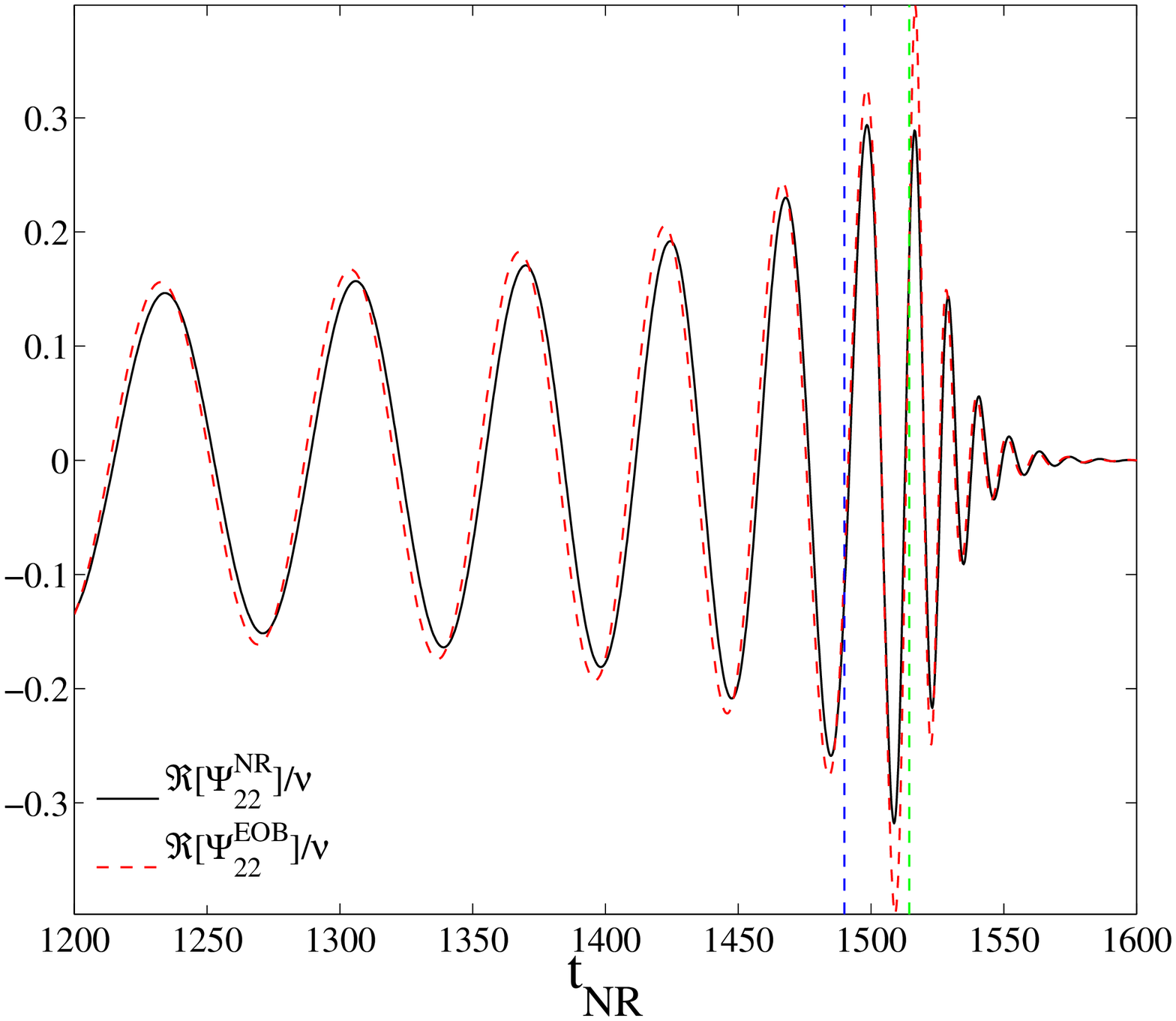} 
  \end{center}
\caption{\label{label:fig3} Comparison between the EOB restricted waveform 
approximation, Eq.~(\ref{psiEOB_BD}), and NR for $a_5=60$ and 
$v_{\pole}=v_{\rm pole}^{\rm DIS}(\nu=1/4)=0.6907$: 
 frequencies (top--left), phase difference
   (top--right), amplitudes (bottom--left) and real parts (bottom--right) 
   of the two gravitational waveforms. The vertical line at $t_{\NR}=1510$
   locates the maximum of (twice) the orbital frequency $\Omega$ 
   (alias the ``EOB-light-ring'') and indicates the matching time.  
   The vertical dashed line at $t^{\NR}=1490$ indicates the crossing 
   time of the adiabatic LSO orbital frequency ($\Omega_{\rm LSO}= 0.1081$).} 
\end{figure*}

Let us now consider our second representative value of the effective
4~PN radial potential parameter, $a_5 = 60$. As before we chose 
$\delta = 1.7 M_{\rm f}$. We also selected the same phase ``pinching'' 
interval as above. Then, by trial and error, we found that 
$v_{\rm pole}=0.5356$  yields an excellent EOB/NR 
agreement
\footnote{Note that this ``best'' value of $v_{\pole}$ 
(for $a_5=60$ and $\nu=1/4$) happens to be numerically close 
to the best fitting $v_{\pole}\simeq 0.53$ value that 
Ref.~\cite{arXiv:0711.2628} found in the test-mass 
limit $\nu\to 0$.}.

We exhibit our results in the four panels of Fig.~\ref{label:fig2},
which are entirely parallel to those of Fig.~\ref{label:fig1}. 
The remarkable level of EOB/NR agreement that we get now, when $a_5=60$,
is rather close to the one that we got above when $a_5=25$. 
At this stage, there is no rationale for saying that either value of
$a_5$ is preferred over the other (though $a_5=25$ yields 
somewhat better results). Some partial numerical tests that 
we performed suggest that this conclusion extends to (at least) all 
values of $a_5$ between $25$ and $60$.   

Some of the numbers quantifying the EOB/NR agreement are:

(i) the (two-sided)  EOB-NR phase difference over the time
interval $ (500 M, 1550 M)$ (which covers about 13 GW cycles 
of inspiral, plunge, and most of the ring-down) is smaller than 
about $ \pm \frac{1}{2} 0.13$ radians, which corresponds 
to $\pm 0.01$ GW cycles;

(ii) during the interval $(1100M,1400M)$ the fractional 
EOB-NR amplitude difference varies between $-0.8\%$ and $+0.55\%$.
After $t_{\NR}=1400M$, this fractional difference increases 
from $+0.55\%$ to a maximum of $+23\%$ 
(reached at $t_{\NR}\simeq 1511M$) and then decreases to 
take values of order $+6\%$ during the observationally 
relevant part of the ringdown.

\subsection{Contrasting resummed EOB with restricted EOB, 
for $a_5 = 60$, by  comparing NR to a standard restricted EOB
waveform}
\label{sbsc:restricted}

Finally, we wish to illustrate the importance (for reaching a high 
level of accuracy) of the various ingredients used in our present, 
resummed version of EOB (using a time-extended ``comb matching'' to 5 QNMs) by 
comparing NR to the type of simpler implementation of the EOB framework
used in~\cite{arXiv:0706.3732}. Using again $a_5 = 60$ (which was chosen
as best value in~\cite{arXiv:0706.3732}), we compare NR to the following
implementation of EOB: 

\begin{itemize}
\item we use for $v_{\rm pole}$  the ``standard'' value $v^{\DIS}_{\rm pole}(\nu)$ 
advocated in~\cite{Damour:1997ub}.
\item  we use (as originally proposed in Ref.~\cite{Buonanno:2000ef}) 
the following (Newtonian-order and Kepler-law-assuming) restricted 
quadrupole waveform 
\begin{equation}
\label{psiEOB_BD}
\Psi^{\rm NK}_{22}(t) = -4\nu\sqrt{\dfrac{\pi}{30}} \Omega^{2/3} \exp(-2\i\Phi) \ ,
\end{equation}
without any explicit  PN  ($F_{22}$) corrections, nor any NQC ($a,b$) corrections.

\item we use only $3$ (positive-frequency) QNMs.
 
\item and, we match the plunge
and ring-down waveforms in a very small interval ($\delta/M_{\rm f} = 0.2$ instead of our
preferred $1.7$) around the maximum of the orbital frequency. 
[Indeed, the  matching of the two waveforms and their derivatives at 
a sharply defined moment is equivalent to considering the $\delta \to 0$ 
limit of our comb-matching technique]. 
 
\end{itemize}

The results of such a coarser EOB implementation are shown in 
Fig.~\ref{label:fig3} (which is parallel to the previous two 
figures). By contrasting Fig.~\ref{label:fig3} with Fig.~\ref{label:fig2} 
(which used the {\it same value} of $a_5$), we see that: 

\begin{itemize}

\item the EOB frequency agrees less well with the NR one than before,
especially around the matching point. Note in particular that 
the post-matching analytical frequency jumps up from the
maximum (doubled) orbital frequency significantly more vertically than before,
thereby decoupling too soon from the exact frequency, and accruing a larger dephasing
than before (because of the too localized matching, and -- to a lesser degree --
the use of only 3 QNMs).

\item the EOB-NR (maximal) phase difference over the same time
interval $ (500 M, 1524 M)$ is about 2.2~times larger than before.
One now ends up with a phase difference of $\pm \frac{1}{2}0.29$ radians, 
i.e. $0.023$ GW cycles over about 13~GW cycles. 
The top-right panel of Fig.~\ref{label:fig3} illustrates the 
fact that matching with 5~QNMs (dashed line) reduces the 
dephasing accumulated during the transition from merger to ringdown.

\item the modulus of the analytical waveform is now distinctly
larger than the NR one during the inspiral (because of the lack of PN corrections).

\item the modulus also exhibits a more significant discrepancy ($+35\%$) 
with the NR one at the end of the plunge (because of the use of the 
Kepler-law-assuming $\propto \Omega^{2/3}$, which, as pointed out 
in~\cite{Damour:2006tr}, tends to overestimate the amplitude).
 
\item Note also that one visually notices these differences at the level
of the GW waveforms.

\item The same resummed-EOB/restricted-EOB comparison was done 
      in~\cite{arXiv:0705.2519}, in the $\nu\ll 1$ case, with similar
      conclusions.

\end{itemize}  
In spite of these relative blemishes,  note, however that 
this ``coarser'' EOB-type implementation still succeeds in following
the phase of the exact signal to $\pm 0.023$ GW cycles over about 
13~GW cycles.

Note that the corresponding EOB/NR agreement exhibited in Fig.~4 of 
Ref.~\cite{arXiv:0706.3732} seems to be somewhat better\footnote{The reader should however 
keep in mind that in Fig.~4 of Ref.~\cite{arXiv:0706.3732} the EOB-NR
phase difference is divided by $2\pi$ compared to the one showed in 
our Fig.~\ref{label:fig3}.} than the one exhibited by our
Fig.~\ref{label:fig3}. This difference might have several origins,
notably: (i) a difference in the accuracy of the NR data\footnote{
The data used in~\cite{arXiv:0706.3732} did not benefit from 
the reduction in eccentricity used in the data considered here.},
and (ii) a difference in the procedure used to best shift time and
phase between EOB and NR data.

\section{Conclusions}
\label{sec:conclusions}

We have compared  a recently proposed,  resummed  $3^{+2}$-PN accurate
Effective-One-Body (EOB) waveform to the result of a numerical simulation
of a coalescing equal-mass binary black hole performed at
the Albert Einstein Institute. We find a remarkable
agreement, both in phase and in amplitude, between the new EOB waveform and
the  numerical data. More precisely, we find that the maximal dephasing 
between EOB and numerical relativity (NR) can  be reduced below $\pm 0.005$ 
GW cycles over the last $\sim 900 M$ (corresponding to about 12~GW cycles 
plus ringdown ones) of the simulation. This level of agreement was 
exhibited for two representative values of the effective 4~PN parameter 
$a_5$, namely $a_5 = 25$ and $a_5 = 60$, and for a corresponding, 
appropriately ``flexed'' value of the radiation-reaction resummation 
parameter $v_{\rm pole}$. In addition, our resummed EOB amplitude agrees 
to better than the $1\%$ level with the NR one
up to the late inspiral.
  
We have also compared the NR data to a coarser implementation of the
EOB approach (restricted waveform, standard $v^{\rm DIS}_{\pole}$, 
instantaneous matching to 3 QNMs). The EOB/NR agreement is slightly
less good in this case, though the phase agreement remains quite good 
($\pm 0.023$ GW cycles over the last $\sim 1000 M$ of the simulation).

Let us point out a notable feature of our results. In the recent work of
Damour and Nagar~\cite{arXiv:0711.2628}, the same resummed $3^{+2}$-PN accurate
EOB waveform was compared to a long, very accurate equal-mass inspiral
simulation of the Caltech-Cornell group~\cite{arXiv:0710.0158}. 
It was found that an excellent EOB/NR agreement was obtained when 
$a_5$ and $v_{\rm pole}$ were following the rather precise correlation 
plotted in the upper panel of Fig.~3 of Ref.~\cite{arXiv:0711.2628}. 
Let us denote this correlation as $ a_5 \to v_{\rm pole}^{\rm best\,inspiral}(a_5)$.
In the present paper, we similarly found that the EOB/NR
agreement was at its best when, for a given $a_5$,\footnote{Though we did not
explore all possible values of $a_5$, we sampled intermediate values between
the representative $a_5$ values we picked and convinced ourselves that the
same conclusion held for them.} $v_{\pole}$ was taking a rather precise
corresponding ``best fit value'', say $v_{\pole}^{\rm best\,insplunge}(a_5)$. 
In particular, we found $v_{\rm pole}^{\rm best\,insplunge}(25) = 0.6241$
and $v_{\rm pole}^{\rm best\,insplunge}(60)= 0.5356$. 
On the other hand, the results of ~\cite{arXiv:0711.2628} yield 
$ v_{\rm pole}^{\rm best\,inspiral}(25) = 0.5340$, and 
$ v_{\rm pole}^{\rm best\,inspiral}(60) = 0.4856$. 
The differences between these sets of values are 
$v_{\rm pole}^{\rm best\,insplunge}(25) - v_{\rm pole}^{\rm best\,inspiral}(25)
= 0.0901$ and 
$v_{\rm pole}^{\rm best\,insplunge}(60) - v_{\rm pole}^{\rm best\,inspiral}(60)
= 0.0500$. Note also that the ``best insplunge'' $v_{\rm pole}$ values are in between
the ``best inspiral'' ones and the originally advocated~\cite{Damour:1997ub} one 
$v^{\rm DIS}_{\rm pole}(\nu = 1/4)= 0.6907$. This finding will deserve further investigation
in the future. At this stage we can only speculate on the various possible origins
of this difference: (i) it might be due to the fact that, not having access to the
original NR data of~\cite{arXiv:0710.0158}, Damour and Nagar had to rely on
rather coarse measurements extracted from published figures; (ii) it might be due
to systematic errors in the NR data of~\cite{arXiv:0710.0158}; (iii) it might
alternatively come from systematic errors in the NR data used in the present
paper; (iv) it might come from the fact that the ``best-fit'' 
${\cal F}_{\varphi}(v_{\rm pole})$ is not a uniform approximation  
(as a function of frequency) to the exact radiation reaction 
(see, in the  $\nu\to0$ limit, the bottom panels of Fig.~1 in~\cite{arXiv:0711.2628}) 
and, finally, (v) it might come from some ``missing physics'' in the resummed
EOB waveform explored here. There are several candidates for this missing physics.
One suggestion (which follows the original suggestion of~\cite{gr-qc/0103018})
is that one might need to consider still higher (uncalculated) PN contributions
to the radial EOB  potential\footnote{For simplicity, we consider only
linear-in-$\nu$ higher PN contributions. If the need arises (and the fact that the
unequal-mass EOB/NR comparisons of~\cite{arXiv:0706.3732} seem to exhibit a
strong dependence on the mass ratio might suggest it) one can easily add in
a non-linear $\nu$ dependence.} $A(u) = 1 - 2 u + 2 \nu u^3 + a_4 \nu u^4
+ a_5 \nu u^5 +a_6 \nu u^6 + \cdots$ where $u=1/r$. 
Another suggestion is that non-quasi-circular (NQC) corrections to radiation reaction 
might modify the phasing during  late inspiral and plunge. As an
example, we have looked at this possibility. More precisely, following~\cite{arXiv:0705.2519},
we can introduce a new flexibility parameter $\bar{a}^{\rm RR}$
\footnote{Actually~\cite{arXiv:0705.2519} introduced a parameter ${a}^{\rm RR}$
which is, roughly, the negative of $\bar{a}^{\rm RR}$, with a NQC
radiation reaction factor of the form $1+ a^{\rm RR}p_{r_*}^2/(r\Omega)^2$ }
such that the radiation reaction force is multiplied by a
correction factor $f_{\rm RR}^{\rm NQC}$ given by

\be 
\label{rr_nqc}
f_{\rm RR}^{\rm NQC} = 
\left( 1+ \bar{a}^{\rm RR}\dfrac{p_{r_*}^2}{(r\Omega)^2 + \epsilon} \right)^{-1}.
\ee
Such a factor will be very close to one during the inspiral (and therefore
will be negligible in the EOB comparison to the Caltech-Cornell data), but will
start being significantly less than one (if $\bar{a}^{\rm RR} > 0$)
during the late inspiral and plunge, which are of
interest for the comparison to the presently considered data.
And indeed, we have found that by choosing a value $\bar{a}^{\rm RR} \sim + 40$
(and $\epsilon = 0.12$ as in the waveform NQC factor considered above)
we could, when $a_5 = 25$, obtain an excellent EOB/NR fit by using the 
``best inspiral'' value $  v_{\rm pole}^{\rm best\,inspiral}(25)
= 0.5340$  (instead of the above $v_{\rm pole}^{\rm best\,insplunge}(25) = 0.6241$).
This issue needs to be further investigated by using the most accurate possible data 
covering both inspiral and plunge. We hope to come back to it in the future.

Finally, we think that the present work, taken in conjunction with other 
recent works on the EOB/NR comparison, confirms the 
ability of the EOB formalism to accurately capture the  general-relativistic 
waveforms. The present work has also shown that the recently proposed resummed 
$3^{+2}$-PN accurate waveform is important for  defining analytical EOB 
waveforms that faithfully represent (both in phase and in amplitude) the 
waveforms emitted by equal-mass coalescing (non-spinning) black hole binaries.

\acknowledgments We thank Peter Diener for assistance and discussion
in the early stages of this work, Sascha Husa for help with the NR
initial data, and Sebastiano Bernuzzi for help in the analysis of the
ringdown waveform. The NR computations were performed with the
Damiana, Belladonna and Peyote clusters of the Albert Einstein
Institute. This work was supported in part by DFG grant
SFB/Transregio~7 ``Gravitational Wave Astronomy''. The activity of 
AN at IHES is supported by INFN. The commercial software products 
Mathematica and Matlab have been extensively used in the preparation 
of this paper.

\end{document}